\documentclass[prd,preprint,a4paper,showpacs,showkeys]{revtex4-1} 

\usepackage{bm}
\usepackage{amsmath}
\usepackage{amssymb}
\usepackage{graphicx}
\usepackage{color}

\def\p{\partial}
\def\a{\alpha}
\def\b{\beta}
\def\d{\delta}

\def\ve{\varepsilon}
\def\k{\kappa}
\def\l{\lambda}
\def\L{\Lambda}

\def\G{\Gamma}

\def\O{\Omega}

\def\ra{\rightarrow}

\def\Mfunction#1{\mathop{\rm #1}\nolimits}

\begin{document}

\title{New renormalization group study of the 3-state Potts model
and related statistical models}

\author{Jos\'e Gaite}
\affiliation{
Applied Physics Dept.,
ETSIAE,
Universidad Polit\'ecnica de Madrid,\\ E-28040 Madrid, Spain}

\date{July 22, 2024}

\begin{abstract}
The critical behavior of three-state statistical models invariant under the full symmetry 
group $S_3$ and its dependence on space dimension have been a matter of interest and debate. 
In particular, the phase transition of the 3-state Potts model in three dimensions is believed to be of the first order, without a definitive proof of absence of scale invariance in three-dimensional field theory with $S_3$ symmetry. This scale invariance should appear as a non-trivial fixed point of the renormalization group, which has not been found.  
Our new search, with the non-perturbative renormalization group, finds such a fixed point, 
as a bifurcation from the trivial fixed point at the critical space dimension $d=10/3$, 
which extends continuously to $d=3$. It does not correspond to 
a second-order phase transition of the 3-state Potts model, but is interesting in 
its own right. In particular, it shows how the $\ve$-expansion can fail.
\end{abstract}


\maketitle

\section{Introduction}
\label{intro}

The application of renormalization group techniques produced a breakthrough in the theory of critical phenomena in statistical physics. While the Landau theory of critical phenomena 
already emphasized the fundamental role of symmetry in second-order phase transitions
\cite{LL}, 
renormalization group theory allowed the classification of universality classes 
according to symmetry and space dimension. The space dimension is important, 
because, in high dimensions, 
a mean-field approach suffices to describe the possible phases and their symmetries, 
whereas, in low dimensions, fluctuations predominate and critical phenomena are richer and more complex. 
This dimension dependence is observed in the simple and much studied case of 
the reflection symmetry group $\mathbb{Z}_2$, in which one does not have non-trivial critical phenomena for dimensions
$d>4$, while there exists one non-trivial type of critical phenomena in 
three dimensions, given by the Wilson-Fisher renormalization group fixed point, and there exist many types in two dimensions.

Naturally, other finite groups give rise to richer critical phenomena. 
The usual generalization of the $\mathbb{Z}_2$ symmetry is either the $\mathbb{Z}_q$ 
symmetry 
or the full symmetry group $S_q$ of permutations of $q$ states, which gives rise to the family of Potts models (models with $\mathbb{Z}_q$ symmetry are sometimes called 
planar Potts models).
These symmetries are less studied.  
Regarding the Potts models, it is interesting to consult Wu's 1982 review 
\cite{Wu82}, 
where he lists several then unsolved problems; in particular, the formulation of 
``a rigorous argument proving or disproving that the $d=3$, $q=3$ transition is first order.'' 
This is the kind of problem that can be addressed 
by means of renormalization group techniques. However, the problem is not totally solved, 
in spite of considerable advances towards a solution. 

The Potts model Landau potential is easily constructed in terms of a mean-field 
approximation \cite{Wu82}. A convenient Hubbard-Stratonovich transformation of its 
partition function readily provides a potential in terms of a multi-component field, which 
encompasses all the models with $S_q$ symmetry
\cite{ZW75}. For $q=2$, namely, for the $\mathbb{Z}_2$ universality class, 
we have the standard potential of a single field, symmetric under the change of the field 
sign, and therefore with only even powers of the field. 
For $q>2$, the symmetry allows a term cubic 
in the field \cite{Wu82,ZW75}. However, according to the Landau theory, phase transitions of the second kind can exist only if the third-order terms in 
the expansion of the potential vanish identically \cite{LL}. Nevertheless, the field
fluctuations could render this argument invalid 
\cite{S-Fisher,Alex74}. It is known that, for $q>4$ and $d\geq 2$, 
fluctuations are insufficient for that purpose, 
but the phase transition of the 3-state Potts model is definitely of the second 
order in two dimensions and open to question in three dimensions \cite{Wu82}.	
Models with $S_q$ symmetry and multisite interactions \cite{Wu82} are also 
interesting and are naturally included in the field-theory formulation 
\cite{ZW75}.

The renormalization group (RG) is the adequate tool to analyze the effect of 
field fluctuations on the effective potential of a field theory. The RG has been employed 
for the Potts model since Golner's early investigation \cite{Golner73}, which 
concluded that the phase transition of the 3-state Potts model in $d=3$ is of the first order.
Golner employed Wilson's recursion formula and, actually, his work was 
simultaneous with Wilson and Kogut's writing 
the celebrated report ``The renormalization group and the $\ve$ expansion'' 
\cite{Wil-Kog}.  
Afterwards, renormalization group studies have mostly employed perturbation theory and the 
$\ve$-expansion \cite{ZW75,Amit76,mcrit-LP} or 
Wilson's exact renormalization group itself \cite{Newman84,MOP,BAZ,S-V} (Golner \cite{Golner73} already mentions the possibility of using the exact renormalization group). 

Zia and Wallace employed the $\ve$-expansion \cite{ZW75}, and chiefly considered the 
isotropic case, that is to say, 
the standard Wilson-Fisher RG fixed point, or else the case of ``cubic anisotropy''  
due to a quartic field term. 
They made some comments on the qualitative effect of the symmetry breaking produced 
by the cubic field term, associated to the $S_3$ symmetry.
Amit \cite{Amit76} pointed out that the upper critical dimension that corresponds to the 
cubic field term is $d=6$ and considered an $\ve$-expansion about it.
Newman et al's exact renormalization group formulation \cite{Newman84} 
also appealed to $\ve$-expansion arguments,  
combining the $\ve$-expansions about $d=4$ and $d=6$. 
These complications can be avoided by working in a fixed dimension. 
For example, Fucito and Parisi \cite{FucitoP} worked out a one-loop perturbation theory
in arbitrary dimension but kept this dimension fixed.
(In the appendix, {sect.~\ref{Perturbative},} 
we briefly describe the one-loop renormalization of a more general theory in $d=3$.) 

The exact renormalization group (ERG) underwent extensive developments along several decades 
\cite{Wil-Kog,Wegner-H,N-C-Stanley,N-C-Stanley_1,Polchinski,Hasen2,Felder,MOP,Wett,Alford,Morris,Morris_1,Morris_2}. 
Margaritis, \'Odor and Patk\'os \cite{MOP} applied the Wegner-Houghton formulation of 
the ERG for the vector model \cite{Wegner-H} to the case with $S_3$ anisotropy,  
which they connected with the 3-state Potts model. 
From the absence of RG fixed points with a single stable direction, 
they concluded that its phase transition is ``unambiguously first order''. Thus, 
Wu's unsolved problem should have been settled. 
However, the argument of Margaritis et al \cite{MOP} is questionable.  
In fact, they employed a truncation of the Wegner-Houghton ERG 
which does not really provide the ``rigorous argument'' that Wu had called for. 
We shall expound on Margaritis et al's argument later.

The exact renormalization group is currently a mature and active area of research, with several recent reviews \cite{KopietzBS,Delamotte,Rosten,Dupuis}. Moreover, 
several researchers are still working on the treatment of the Potts model within the ERG approach \cite{BAZ,S-V,Wiese}. 
Ben Ali Zinati and Codello's extensive study \cite{BAZ} draws interesting conclusions 
in high $d$ but postpones a full study of $d=3$.  
The problem that they encounter is the need of 
too many coupling parameters to have a convergent approach in $d=3$. 
Another paper of the same group \cite{Plato} combines 
a ``functional perturbative renormalization group''
with the $\ve$-expansion, in a manner that resembles the work of Newman et al 
\cite{Newman84} but that is more general, given that they consider a larger set of 
upper critical dimensions. In particular, they consider $d=10/3$, 
a dimension that is only noted by Newman et al as a ``coincidence'' with $d=3.4$, this being
the lower dimension at which their scheme seems to fail. 
The dimension $d=10/3$ will feature prominently in our own analysis.

The role of $d=10/3$ had been already studied by Codello, Safari, Vacca and Zanusso 
\cite{Blume-Capel},  
in the single-field theory case, with a quintic potential
(which they named Blume-Capel universality class). 
They employed a combination of perturbation theory and the $\ve$-expansion. 
The $\ve$-expansion for potentials with leading odd terms was actually considered long ago 
by Nicoll, Chang, and Stanley \cite{N-C-Stanley} and 
it has also been studied recently by Gracey \cite{Gracey}. 
Ben Ali Zinati et al associate $d=10/3$ with pentagonal symmetry in Ref.~\onlinecite{Plato}. 
Codello, Safari, Vacca and Zanusso \cite{mcrit-LP} have further studied 
the role of $d=10/3$ as an upper critical dimension for the Potts model,  
and they briefly consider the case $q=3$.

Several of the previous works, namely, Refs.~\onlinecite{Newman84,BAZ,mcrit-LP,S-V,Wiese},
consider the $q$-state Potts models for general $q$ and, furthermore, 
assume that $q$ can vary continuously. Here, we focus on $q=3$ and, in general, 
on the anisotropy of the vector model due to a cubic field term. 
On the other hand, 
our method is based on the sharp-cutoff Wegner-Houghton ERG, unlike some 
recent methods
\cite{BAZ,S-V}. 
We have two reasons for that choice. 
The first one is that it gives rise to simpler algebraic equations, namely, the ones 
first derived by Margaritis, \'Odor and Patk\'os \cite{MOP}. 
The second reason is that we have found 
recently that the simple sharp-cutoff regularization works as well as others 
when we consider broad regions of the parameter space 
\cite{II}. 
{Nevertheless, we briefly study the ERG equations in another scheme.} 

Like Ben Ali Zinati and Codello \cite{BAZ}, we find that a clearly convergent scheme is not easy to achieve. Indeed, the 
search for fixed points of the ERG
leads to complicated algebraic calculations, which are eventually 
beyond our present computing capabilities. 
We find that the most promising approach seems to consist of 
a combination of the exact renormalization group with the study of 
bifurcations from the trivial solution, like in the $\ve$-expansion.  
In fact, our approach 
considers fixed $q=3$ and a bifurcation at 
$d=10/3$. In general, it is necessary to 
consider the various bifurcations from the trivial RG fixed point that appear at 
dimensions $d \leq 6$. Our approach requires tools from the theory of bifurcations 
of dynamical systems and, in this regard, connects with Gukov's work \cite{Gukov}.

The detailed study of the bifurcation at $d=10/3$ allows us to actually find a 
non-trivial RG fixed point with $S_3$ symmetry in $d=3$. 
However, we discover that this fixed point is beyond the reach of the $\ve$-expansion 
based on $d=10/3$. 
{Therefore, the fixed point could not have been found with previous approaches based on the 
$\ve$-expansion}
\cite{Newman84,BAZ,mcrit-LP,S-V}. 
In fact, to establish the physical meaning of the new fixed point, we need to consider 
global properties of the RG flow. 
The study of global features of nonlinear renormalization-group equations 
was begun by Nicoll, Chang, and Stanley \cite{N-C-Stanley_1}. 
Here, we face a much more complex set of nonlinear equations. 
and we leave a comprehensive study for the future. 

Finally, 
let us emphasize that field-theory methods apply to universality classes of definite symmetry but do not pinpoint any statistical model. Among models 
with $S_3$ anisotropy, the 3-state Potts model is the simplest one, 
but we must consider other models, with parameters other than temperature. 
Surely, the next $S_3$-model to consider is the diluted 3-state Potts model \cite{Wu82}. 
It has two parameters, namely, temperature and concentration. 
This model turns out to be useful to study how critical properties of the Potts model 
change with $q$ and $d$ \cite{Nien-RS}.
Of course, generalizations of the 3-state Potts model with more than two parameters
have been devised, and we shall mention some pertinent ones. 

{The main text consists of seven main sections and a concluding section. At the end, 
there are two appendices: the first one points to some relevant results of 
perturbation theory and the second one shows a few ERG beta-functions with different regulators.} 
Section \ref{MOP_t} summarizes Margaritis et al's results in Ref.~\onlinecite{MOP}, 
connects them with later work, and adds a little extension of those results. 
Section \ref{Phd} places the 3-state Potts model in the context of general $S_3$-models, 
in arbitrary dimension but with special regard to two dimensions, where a wealth 
of information on critical phenomena is available.
Section \ref{ERG-ee} examines the possible 
bifurcations from the trivial fixed point and what the respective $\ve$-expansions 
are like. 
Dimension $d=10/3$ is singled out, but the $\ve$-expansion based on it is not useful. 
Section \ref{S3FP} shows that the bifurcation at $d=10/3$ can nonetheless be studied with 
fully non-perturbative methods, which leads to the discovery of an $S_3$-invariant 
field theory in three dimensions. 
{Section \ref{Litim} is concerned with the problem of regularization dependence of 
the results.} 
Section \ref{RG-FP} tries to extract some information of the nonlinear renormalization-group equations from their linearization and proposes how to connect with 
fixed-dimension perturbation theory. 
A summary and some further ideas appear in the concluding section.

\section{
Truncations of the Wegner-Houghton equation}
\label{MOP_t}

Margaritis, \'Odor and Patk\'os \cite{MOP} applied the Wegner-Houghton formulation of 
the ERG \cite{Wegner-H} to the $N$-component vector model, either with or without a cubic field term, 
employing suitable truncations of the coupling constant space to derive manageable systems of 
ordinary differential equations, that is to say, equations that can be treated 
numerically. 

The Wegner-Houghton sharp-cutoff ERG equation, in the local potential approximation,  
expresses how the effective potential $U$ changes when the cutoff $\L$ shrinks. 
For the $N$-vector model, it is a nonlinear partial differential equation that is written as 
\begin{equation}
\frac{\p U(\phi_\a,\L)}{\p \L} = -\frac{A_d}{2} \,\L^{d-1}\,
\ln\det\left[\L^2 \d_{\a\b}+ U_{\a\b}(\phi_\a,\L)\right],
\label{ERG}
\end{equation}
where $\a,\b=1,\ldots,N$ and $A_d = \O_d/(2\pi)^d$ 
($\O_d$ is the area of the $d$-dimensional unit sphere). 
The integration of the ERG equation over $\L$, between an initial cutoff $\L_0$
and the final value zero, carries out the elimination of field fluctuations on all scales 
and transforms the ``bare'' potential at $\L_0$ into a renormalized one.

Margaritis, \'Odor and Patk\'os's \cite{MOP} expanded the effective potential in 
a power series of the fields and then truncated this series to a polynomial. 
The expansion of $U(\phi_\a,\L)$ in powers of $\phi_\a$ defines a 
set of coupling constants $C_i(\L),\,i=1,\ldots,$ etc. Besides, 
Margaritis et al \cite{MOP} made the standard 
redefinition to dimensionless variables, namely, 
$$\phi_\a \ra \phi_\a/\L^{d/2-1},\quad U \ra U/\L^d,\quad \L \ra t=\ln(\L_0/\L).$$
The dimensionless 
coupling constants, generated by the expansion in dimensionless variables,  
are denoted here, as in Ref.~\onlinecite{MOP}, 
by lower case letters $c_i(t),\,i=1,\ldots,$ etc.
We shall employ both dimensionless and dimensionful coupling constants. The relation 
between them can be non-trivial in the limit $\L\ra 0$ ($t\ra \infty$).

As regards the isotropic $O(N)$ vector model, Margaritis, \'Odor and Patk\'os's 
early work  \cite{MOP} preceded later work with the ERG, achieved without a truncation of the 
coupling constant space \cite{Comellas-T}.
However, the agreement between both sets of results is not satisfactory, in spite of the 
fact that Margaritis, \'Odor and Patk\'os kept a number of coupling constants 
that suffices for $N=1$, that is to say, for the single scalar field theory. 
It appears that a growing number of coupling constants is needed for larger $N$, as is 
already manifest in Table~1 and Fig.~1 of Ref.~\cite{MOP}. 

For the two-component field theory with a cubic field term and $S_3$ symmetry, 
Margaritis, \'Odor and Patk\'os \cite{MOP} found that the isotropic (Wilson-Fisher) 
RG fixed point has two relevant directions, one in the isotropy 
subspace and another that breaks isotropy to the $S_3$ symmetry. 
The latter could lead to a $S_3$-invariant RG fixed point, 
but they concluded that such point does not 
exist. They employed a 13-dimensional coupling-constant space in their calculations. 
Of these 13 coupling constants, only 5 belong to the isotropic 
subspace (see below).

We must notice that Table~1 and Fig.~1 of Ref.~\cite{MOP} show that no convergence 
is achieved in the isotropic model with only 5 coupling constants. 
In spite of it, 
Margaritis, \'Odor and Patk\'os considered their truncation of the $S_3$ model 
sufficiently reliable to assert 
that the phase transition of the 3-state Potts model is ``unambiguously first order''. 
Moreover, in order to assert that, they had to sift through the numerous fixed-point  solutions that appear in every truncation and discard the spurious ones.  
This is a task that cannot be carried out with mathematical rigor, unless one 
appeals to other arguments, as will be argued here.

As for convergence of truncations, the simplest test is to calculate within an expanded constant coupling space, that is, to include higher degree couplings in the potential, and then evaluate how much the results change.
Margaritis, \'Odor and Patk\'os's truncation of the $S_3$ potential includes
up to $\phi^{10}$ ($\phi$ being either $\phi_{1}$ or $\phi_{2}$, the two fields for $N=2$) 
\cite{MOP}. 
To study $S_3$ anisotropy, 
it is convenient to write the potential $U$ in terms of the two $S_3$ invariants, namely,
\begin{align}
&I_1=\phi_{1}^{2}+\phi_{2}^{2},\quad I_2=\phi_{1}^{3}-3\phi_{1}\phi_{2}^{2}\,,\nonumber\\
U(I_1,I_2)&= 
{c}_{1}I_1+{c}_{2}I_2+{c}_{3}I_1^{2}+{c}_{4}I_1I_2+{c}_{5}I_1^{3}+{c}_{6}I_2^{2} + 
\nonumber\\
& {c}_{7} I_1^2 I_2 + {c}_{8}I_1^{4} + {c}_{9} I_1 I_2^2 + {c}_{10} I_1^{3} I_2 + 
{c}_{11} I_2^{3} + {c}_{12}  I_1^{5} + {c}_{13}  I_1^{2} I_2^{2}
\,.
\label{13pot}
\end{align}
Of the 13 terms, the 5 $I_2$-independent terms preserve the $O(2)$ 
symmetry. In this subspace, the basic critical exponent is $\nu=0.937$ 
\cite[table~1]{MOP}, whereas the value obtained by Comellas and Travesset 
with no truncation 
is $\nu=0.768$
\cite{Comellas-T}. Moreover, Table~1 and Fig.~1 of Ref.~\cite{MOP} show that there 
are considerable fluctuations in the value of $\nu$ up to the truncation at 
$\phi^{16}$ of the isotropic 
field theory, with 8 terms (the maximum number that 
Margaritis et al consider in that case).

Since potential (\ref{13pot}) 
is insufficient, it behooves us to explore extensions of it. 
Given that the truncation at $\phi^{11}$ is somewhat awkward, we consider 
the truncation at $\phi^{12}$, that is to say, we add to potential (\ref{13pot}) 
the following terms:
\begin{equation}
\d U(I_1,I_2) = {c}_{14} I_1^{4} I_2 + {c}_{15} I_1 I_2^{3} + {c}_{16}  I_1^{6} +
{c}_{17} I_1^{3} I_2^2 + {c}_{18} I_2^{4}\,.
\label{18pot}
\end{equation}
We now have 18 terms altogether, of which only 6 terms are independent of $I_2$. 
Let us compare with Margaritis et al's results to examine what improvement is achieved.

The two positive eigenvalues of the dimension matrix at the 
Wilson-Fisher 
fixed point of potential (\ref{13pot}) calculated by Margaritis et al were 
$\l_1=1.0674$ and $\l_2=1.0362$ (our own calculation confirms these values). 
The most relevant eigenvalue corresponds to the isotropy subspace,  
yielding $\nu=1/\l_1=0.937$, as said above. 
The direction corresponding to $\l_2$ breaks isotropy.
In the 18-dimensional space including (\ref{18pot}),  
we still find two positive eigenvalues for the Wilson-Fisher fixed point, which now are 
$\l_1=1.4766$ and $\l_2=0.8082$, 
the latter isotropy breaking.  
The value $\nu=1/\l_1=0.677$ agrees with Ref.~\cite[table~1]{MOP}, but it is still 
quite different from Comellas and Travesset's above-quoted result  
\cite{Comellas-T}. 
Let us note that the value $\nu=0.677$ is quite close to the ones 
obtained with more refined methods \cite{Jakub}, but it is surely by chance. 

To conclude, 
since even the truncation at $\phi^{16}$ shows no sign of numerical convergence 
in the isotropic subspace \cite{MOP}, 
we need a very large number of terms in the effective potential 
to achieve convergence. Nevertheless, we have a kind of qualitative stability of 
the results, 
which may make us confident that there is no $S_3$ RG fixed point. 
Unfortunately, even this qualitative stability of 
the results of Margaritis et al \cite{MOP} is questionable, 
as they themselves partially admit. 
The problem lies, of course, in the multitude of 
spurious fixed points. The possible number of solutions of 
systems of algebraic equations can be estimated from the B\'ezout bound 
(or also from Khovanskii's bound) \cite{Bezout}. 
The proliferation of roots compounds with another problem: the rapid growth with the level 
of truncation of the numerical coefficients in the equations, 
affecting the precision of the numerical roots.

Therefore, to establish the structure of RG fixed points 
with mathematical rigor, one needs to appeal to other methods. 
Before doing so, let us consider some general facts about 
the 3-state Potts model and other $S_3$ models.

\section{Phase diagram of the 3-state Potts and other $S_3$ models}
\label{Phd}

The 3-state Potts model, in its simplest version, with only a nearest-neighbor interaction, 
has temperature as single parameter, like the Ising model. 
Therefore, its phase diagram in one-dimensional, say, a line. 
In $d=2$, this line only has one remarkable feature, 
a critical point, which can be located thanks to duality relationships 
that exist in some lattices \cite{Wu82}. In $d=3$, a critical point may not exist, 
but one can find another feature in the phase diagram:  
a quadruple point 
\cite{Wu82,S-Fisher}. A quadruple point is such that four phases coexist. In the 
3-state Potts model, such phases correspond to three ordered phases, in which 
one of the three states predominates, and an additional disordered and symmetric phase, 
in which they all are equally represented. This latter phase is, of course, the only one 
that survives at very high temperature, while the three ordered phases are the 
ones present at very low temperature, where the $S_3$ symmetry is spontaneously broken. 

If there is a quadruple point, then 
there must be a range of temperatures in which the four phases are possible, with a variable 
degree of stability. Indeed, taking the quadruple-point temperature, 
a temperature rise makes the ordered phases less stable, as it  
favors the disordered phase. Conversely, a temperature fall makes 
the disordered phase less stable and favors the ordered phases. In a range 
of temperatures, the four phases are stable or metastable and, thus, 
there must be two other remarkable temperatures, 
at which instability of either the order or disorder arises. 

The above picture is substantiated by the mean-field theory of the 
3-state Potts model \cite{Wu82,S-Fisher}. The free energy includes a cubic term 
that should vanish for phase transitions of the second order. 
Therefore, in high spatial dimensions where mean-field theory holds, 
there is no critical point but a first-order phase transition; namely,  
the order-disorder transition at the quadruple point, 
where the free energy of each ordered phase is equal to that of the disordered phase.

Mean-field theory, with the consequent first-order phase transition, holds 
in high dimensions but ceases to be valid in low dimensions, where 
field fluctuations gain importance. For example, we know that the isotropic vector 
model mean-field theory is not valid for $d < 4$, because of 
field fluctuations. 
Field fluctuations 
in the 3-state Potts model renormalize the cubic term of the effective potential and could 
make it vanish 
\cite{S-Fisher,Alex74}.
Of course, the effect of fluctuations is best studied with the renormalization group
(Sect.~\ref{RG} below).

As already remarked, an $S_3$ field theory 
represents many statistical models with that symmetry. Naturally, 
there have been elaborations of the basic 3-state Potts model,  
such as models including vacancies \cite{Wu82,Nien-RS}, 
next-to-nearest-neighbor interactions \cite{Gavai-K}, etc. 
Notably, the addition of ``invisible states'' can change the order of the phase 
transition \cite{Krasnytska}. 
From a phenomenological standpoint, 
the study of 
phase diagrams 
intends to connect with experimental results \cite{MuTa} 
(arguably, ``the potential \ldots describes the phase transformations in 
many classes of materials''). Some of these phenomenological potentials certainly 
describe second-order phase transitions. 
The general study and classification of symmetric potentials as well as the 
corresponding analysis of phase stability are best achieved 
with the powerful methods of {\em singularity} or {\em catastrophe} theory \cite{Kutin,I,IMM}.
This is especially true when terms that are explicitly symmetry breaking are considered \cite{IMM}. 

Catastrophe theory puts the Landau theory of phase transitions \cite{LL} 
on a mathematically rigorous framework. {\em Critical points} of a Landau potential, 
that is to say, points where the differential vanishes, 
define the possible phases or other features, which change according to the  
parameter values.  
A potential is maximally 
degenerate for the parameter values such that it has just one critical point. This 
critical point unfolds as the parameters vary. Catastrophe theory provides a 
classification of types of potentials and their unfoldings, summarized in lists of 
{\em canonical forms}. We do not pursue this subject here, although we briefly 
introduce later some methods of the theory of bifurcations of differential equations, 
which greatly overlap with the methods of catastrophe theory. 
For example, the concept of {\em normal form} 
that we shall employ corresponds to the concept of canonical form in catastrophe theory.

\subsection{Renormalization Group analysis}
\label{RG}

In low dimensions, fluctuation effects
are accounted for by the effective potential $U(\phi_\a,\L)$ in Eq.~(\ref{ERG})  
(once $\L$ is integrated over). 
To simplify the potential, given its symmetry, 
we can choose a definite direction in field space, for example, 
by making $\phi_2=0$, 
so that the terms of the potential are simple powers of $\phi_1$. 
Thus, let us write the full dimensional form of potential (\ref{13pot}) as
\begin{equation}
U(\phi_1) = 
{C}_{1}\phi_1^2+{C}_{2}\phi_1^3+{C}_{3}\phi_1^4+{C}_{4}\phi_1^5+({c}_{5}+{c}_{6})\,\phi_1^6 + 
\cdots
\label{6pot}
\end{equation}
(keeping the same notation for the field and the potential, whether or not they 
are dimensional). 
Here, $C_n(\L)=\L^{(5-n)/2}\,c_n(t),\;n=1,\ldots,4,$ 
while both $c_5$ and $c_6$ do not change (they are multiplied by $\L^0$). 
The phase diagram of the $S_3$ potential corresponding to (\ref{6pot}) 
is studied in Ref.~\onlinecite{MuTa}.

Before the ERG equation (\ref{ERG}) was introduced, 
Golner \cite{Golner73} had employed a sort of ERG, called the 
{\em approximate recursion formula} \cite{Wil-Kog}, 
to find the evolution of a potential with only the 
first three couplings in Eq.~(\ref{6pot}). 
The iteration of the recursion formula is practically equivalent to the integration 
over $\L$ in Eq.~(\ref{ERG}). 
With a choice of initial values of $C_1$, 
$C_2$ and $C_3$, and then tuning $C_1$, Golner did not find a fixed point of the recursion 
but instead found a quadruple point. 
This quadruple point is actually a double point of 
the one-field potential (\ref{6pot}). Potential (\ref{6pot}) has, for 
$C_1>0$, a minimum at $\phi_1=0$ such that $U=0$. The condition that it have another 
minimum such that $U=0$ amounts to an algebraic equation to be satisfied by 
the coupling constants. For the five coupling constants in potential (\ref{6pot}) 
(counting ${c}_{5}+{c}_{6}$ as one), the condition is that a seventh-degree polynomial 
in the coupling constants be null. At this point, it can be useful to 
consider some geometrical aspects of coupling constant space and introduce 
concepts that are further employed in the next sections. 

In coupling constant space, an equation for the coupling constants defines a hypersurface, 
that is to say, a space of one dimension less than the full space. If the equation is 
algebraic, then the hypersurface is called a codimension-one 
{\em algebraic variety}, expressing that it is not a differentiable manifold
everywhere, 
because it has singularities along lower-dimensional algebraic varieties. 
Thus, the quadruple-point condition defines a codimension-one algebraic variety.
Generically, two varieties of 
codimension $k$ and codimension $p$ intersect, in $n$-dimensional space,  
if $k+p \leq n$ 
(otherwise, the set of $k+p$ equations is generically incompatible). 
A line generically intersects a codimension-one algebraic variety and, thus, 
going along the line, one meets the variety, as Golner did tuning $C_1$ 
\cite{Golner73}. 

To simplify further, let us restrict ourselves to the first three coupling constants 
in (\ref{6pot}), as Golner did.
The double-minima condition, namely, $U(\phi_1)=0$ for some $\phi_1 \neq 0$, 
implies that 
$${C}_{2}^2=4\,{C}_{1}{C}_{3}\,.
$$
This equation defines a cone in the three-dimensional space of coupling constants.
A line parallel to the $C_1$ axis intersects the cone at one point, such that 
${C}_{1}$ has the same sign as ${C}_{3}$. This sign must be positive 
(on account of the large-field behavior). 
Note that Golner finds a value $C_1<0$ 
in the space of initial (bare) coupling constants, but  
$C_1$ becomes positive under renormalization, of course \cite{Golner73}. 
Under renormalization, the potential also becomes more complicated, as higher-order 
coupling constants take non-null values. 

The generalization to the five-dimensional space of coupling constants
is conceptually simple but algebraically complicated. 
In general, without focusing on a definite form of the potential, 
the 3-state Potts model with a nearest-neighbor interaction has temperature as its single parameter, so that the coupling constants $C_i$ describe a line parametrized by it, 
and we expect that there is a quadruple point at some temperature. The question of 
the existence of critical behavior is more subtle. 
Given the quadruple-point hypersurface, we may find a critical-point on it 
with one further condition. Indeed, let us consider the distance between 
the two minima such that $U=0$, 
which is given by an algebraic expression in terms of the coupling constants. 
The condition that the two minima merge 
raises the codimension to two.
A line does not generically intersect a codimension-two manifold. 
Apparently, the adequate procedure to seek a critical point is by 
tuning two couplings.

From a different point of view, 
the standard condition for a critical point is a vanishing renormalized mass,  
$m^2=2C_1(\L=0)=0$. 
Taking this condition to the potential (\ref{6pot}), as the renormalized potential, 
we find that $\phi_1=0$ is generically not a minimum but an inflection point, 
and to turn it into a real minimum we further need ${C}_{2}=0$. 
Therefore, a critical point should have (at least) codimension two in the full phase diagram.   
Unless ${C}_{1}=0$ implied ${C}_{2}=0$, of course. 
This implication is a possible interpretation of Alexander's argument, namely, the argument 
that renormalization of the cubic term could make it insignificant compared to 
the quartic term \cite{Alex74}.
The existence of a codimension-one critical-point manifold would amount to 
the existence of an RG fixed point with only one relevant parameter. 
The search for an RG fixed point with $S_3$ symmetry and only one relevant parameter 
precisely was Margaritis, \'Odor and Patk\'os's main objective, but it was unsuccessful 
\cite{MOP}.

Let us briefly discuss some aspects of Alexander's argument. 
Alexander \cite{Alex74} focuses on the change of the potential due to 
angular-field fluctuations, that is to say, fluctuations that preserve 
the field norm ($I_1$). 
In the range of parameters where the three ordered states are present, 
one can indeed study the effect of angular-field fluctuations. 
These fluctuations could depress strongly the barrier of the potential between 
each two minima, perhaps to the extent of giving rise to full $O(2)$ 
symmetry. 
Actually, this has been shown to happen for $\mathbb{Z}_q$ models with 
$q\geq 4$ but not to happen for $\mathbb{Z}_3$ (in $d=3$) \cite{Oshi,L-S-Balents}. 
Therefore, it does not happen for potential (\ref{6pot}).
If it were to happen, the $O(2)$ RG fixed point should have only one relevant parameter,  
which we know is not the case.

The arguments above apply to $d=3$. 
In $d=2$, the critical phenomena of all these models are well studied, 
because the theory of critical phenomena  
has the powerful methods of conformal field theory available \cite[ch 9]{SFT}.
Thus, various properties of critical models in $d=2$ can be derived. The 3-state Potts model 
is related to one of the so-called minimal models of conformal field theory, 
with conformal central charge $c=4/5$. 
In contrast, 
the $O(2)$ model is non-minimal, with $c=1$ \cite[ch 9]{SFT}. 
Further comments on $S_3$-invariant models in $d=2$ are made below.

\subsection{$S_3$-symmetry in other dimensions}
\label{other}

Below $d=3$, the effect of field fluctuations is stronger. Indeed, it is
very strong in two space dimensions and especially in one dimension. 
In fact, in one dimension, disorder reigns and there are no proper phase transitions, 
while the theory of two-dimensional phase transitions is rich and well studied. 
Regarding the standard epsilon expansion from $d=4$ 
for the vector model, 
which will be briefly reviewed in Section \ref{ERG-ee}, 
$d=2$ is the integer dimension below $d=3$ to which one can extend downwards.

For $d \leq 4$, a set of fractional dimensions plays an important role in 
the generalization of the standard epsilon expansion. After the formulation of this 
expansion, it was soon discovered that 
the vector model has nontrivial RG fixed-point bifurcations from the Gaussian point 
at a set of definite dimensions, namely, $d_n=2n/(n-1),\;n=2,3,4,\ldots$ \cite{N-C-Stanley}.
Furthermore, in the single-field case, 
those nontrivial RG fixed points were suitable for extrapolation to non-integer dimensions
$d>2$, 
while it was shown that the corresponding effective potentials have precisely $n$ possible minima ($n$ ordered phases) \cite{Felder}. 
It seemed natural that these RG fixed points correspond, in $d=2$, to an infinite set of 
phase transitions, with a growing number of phases. 

In a separate development, in $d=2$, 
the infinite series of unitary minimal models of conformal field theory 
was discovered, 
a series that describes the multicritical behavior of $\mathbb{Z}_2$-symmetric models  
\cite[ch 9]{SFT}. 
A general and direct connection between 
that series of unitary minimal models and the series of 
effective potentials that give rise to the respective types of multicritical behavior  
was fleshed out by Zamolodchikov \cite{Zamo}. 
He made that connection
by studying how the algebra of {\em composite} fields generated by an effective potential 
can reproduce the field algebra of a conformal field theory model, 
obtained by algebraic methods. 
Finally, there have been found 
dimension-interpolating scaling solutions of the ERG equations for 
$\mathbb{Z}_2$-symmetric potentials that  
converge to the conformal field theory models \cite{Kuby,Codello}. 
  
By analogy, one would like to find a series of $S_3$-symmetric potentials corresponding to  RG fixed points bifurcating from the Gaussian fixed point at a set of definite dimensions. Furthermore, assuming that they could be extended down to $d = 2$, they should converge to a set of two-dimensional conformal field theory models with $S_3$ symmetry. Of course, to achieve this goal, one must first determine the appropriate set of conformal field theory models with $S_3$ symmetry.

Actually, an infinite series of 
$S_3$-invariant models of two-dimensional conformal field theory was 
found by Fateev and Zamolodchikov a few years after the discovery of 
the main series of unitary minimal models of conformal field theory 
\cite{FatZamo}. 
These $S_3$-invariant models possess an infinite-dimensional symmetry additional to the conformal symmetry, the so-called 
$W_3$-symmetry, which determines the set of primary fields \cite{FatZamo}. 
The first and simplest model is related to the previously known $c=4/5$ minimal model. 
It has the same central charge, being the only one of the $W_3$-symmetry series with $c<1$.  
However, it only includes the subset of even primary fields under 
the $\mathbb{Z}_2$ symmetry of minimal models. 
In fact, the $S_3$-covariant subalgebra of fields of 
this model can be derived from that of the $c=4/5$ minimal model through 
the condition of {\em modular invariance} of conformal partition functions. It turns out 
to be the first case of the series of {\em non-diagonal} 
modular-invariant partition functions and is denoted as $(A_4,D_4)$
\cite[ch 9]{SFT}. 
After the examination of the allowed dimensions of primary fields, 
it can be identified with the 3-state Potts model \cite[ch 9]{SFT}. 

The diluted 3-state Potts model \cite{Wu82,Nien-RS} has, in two dimensions, 
a tricritical point that is also described by a 
non-diagonal modular-invariant partition function of conformal field theory, namely, 
the one denoted as $(D_4,A_6)$ \cite{Bal-Den,Muss}. This conformal model has $c=6/7<1$ and 
does not belong to the $W_3$-symmetry series.   
The diluted 3-state Potts model has two parameters, namely, temperature and concentration
(one more parameter than the ordinary Potts model). At the tricritical point, 
these two parameters are the coupling constants 
for two conformal fields, which are usually 
denoted as $\epsilon$ and $\epsilon'$ and belong to an 
$S_3$-invariant subalgebra, named thermal algebra \cite{Bal-Den,Muss}. 
The model does not have more $S_3$-invariant conformal fields that, being RG-relevant, 
could be related to couplings not present in the original definition of the model. 

The change of the critical properties of the 3-state Potts model as the dimension varies has 
been widely studied, in both its basic and diluted formulations, since long ago 
\cite{Nien-RS} and until now \cite{Jan-Vila,Chester-Su,Wiese}. 
The consensus is that the critical (or tricritical) behavior of these models that is present 
in $d=2$ extends up to non-integer $d>2$ but ceases below $d=3$.  Of course, more elaborate 
types of $S_3$-symmetric critical behavior in $d=2$ can be suitable for extrapolation up 
to $d=3$. The next more elaborate model to consider is the second model of the $W_3$-symmetry 
series, with $c=6/5$. To analyze this model, 
it is very helpful to consider 
the connection of conformal field algebras with effective potentials, 
initially made for $\mathbb{Z}_2$-symmetric models \cite{Zamo}.
The analogous task for the $W_3$-symmetry series was already 
suggested by Fateev and Zamolodchikov in the original article \cite{FatZamo}, 
and was carried out a few years later \cite{Koh-Yang,IW,IW1}. 

The effective potential for the second model of the $W_3$-symmetry series is read from 
\cite[Eq. 18]{IW} (the case with $p=5$). It can be matched to potential (\ref{13pot}), 
namely, to the truncation at 
$I_1^3$.
This model has three $S_3$-invariant conformal fields that are RG-relevant, which 
can be identified as powers of the elementary fields \cite{IW} and 
match the first three terms of potential (\ref{13pot}), namely, 
the ones before the first two RG-irrelevant terms $I_1I_2$ and $I_1^3$ 
\cite[Eq. 18]{IW}. 
The three RG-relevant conformal fields include two fields of the thermal subalgebra 
($\epsilon$ and $\epsilon'$), 
also present in the tricritical 3-state Potts model, and an extra field, 
equivalent to $I_2$. Of course, there are other RG-relevant fields, which are not 
$S_3$ invariant. Perturbations along all these fields
give rise to the full unfolding of minima of the effective potential and to a complex 
phase diagram, which can be studied with catastrophe theory methods \cite{IW1}.
This is a suitable model to be extended up to $d>2$ and, perhaps, up to $d=3$, 
but this possibility has not been explored yet. 

The above insights from two-dimensional conformal field theory are very interesting, 
but field theory in two dimensions is very particular and upward dimensional extrapolation 
is not trivial. 
In contrast, we have the promising success of dimensional extrapolation down to $d=2$ 
of scaling solutions for $\mathbb{Z}_2$-symmetric potentials bifurcating from the Gaussian solution. 
Of course, the general method of downward extrapolation from the Gaussian solution that 
employs the $\ve$-expansion is much studied and must be considered. 

\section{exact renormalization group and $\ve$-expansions}
\label{ERG-ee}

The $\ve$-expansion has been combined with the ERG since the beginning 
\cite{Wil-Kog,Wegner-H,N-C-Stanley,Newman84,S-V}. 
As dimension at which to expand,
Newman et al \cite{Newman84} considered both 
$d=6$ (the naive upper critical dimension) and $d=4$, but they decided for the latter.  
In so doing, they noticed, in the expansion in $\ve=4-d$, that 
``for the Potts fixed point only the isotropic fields are
of order $\ve$, while the Potts fields are of order 1.'' Therefore, they made 
$q$ a continuous variable and had to 
rely on a critical $q_\mathrm{c}=2$ (the Ising model) and thus consider 
a further expansion in powers of $\d=q-2$, in combination with numerical techniques. 
Variations of this procedure have been proposed recently \cite{S-V,Wiese}. 

The pattern of bifurcations of the isotropic $N$-vector model from the Gaussian solution 
is well understood \cite{N-C-Stanley}. It is independent of $N$.
In the case $N=2$, 
$S_3$ anisotropy occurs if some of the coupling constants that affect $I_2$ are 
non-vanishing in potential (\ref{13pot}) (or extensions thereof). 
The presence of the third-degree term $c_2I_2$ suggests 
an upper critical dimension $d_\mathrm{c}=6$, and hence the corresponding $\ve$-expansion, 
but it may not be the best option \cite{Newman84,S-V}. Actually, 
for $N=1$, $d_\mathrm{c}=6$ was found to be connected with the Lee-Yang edge singularity problem
\cite{Fisher}. 
In addition to the bifurcation at $d_\mathrm{c}=6$, there is another bifurcation, at 
$d_\mathrm{c}=10/3$, such that $c_2$ is non-null, 
due to the term with a fifth power of the field, namely, the term $c_4I_1I_2$.

The need for the term $c_4I_1I_2$, when $c_2\neq 0$ in $d=3$, is also seen 
in a perturbative treatment. Indeed, 
while $c_4$ is irrelevant above $d_\mathrm{c}=10/3$, it is required below 
for renormalizability and, in particular, 
in $d=3$, whenever $c_2\neq 0$. This is evident from the one-loop calculation (in the appendix), which shows that, 
even if the bare coupling constant $C_4$ is null, the presence of the third power of 
the field produces a non-null renormalized value ${C_\mathrm{r}}_4$. 

The role of $d=10/3$ as an upper critical dimension 
in the single-field theory case with a quintic potential
has been studied by Codello et al \cite{Blume-Capel} and by Gracey \cite{Gracey} 
(the single-field theory case is the one considered in the appendix). 
Codello, Safari, Vacca and Zanusso \cite{mcrit-LP} have studied it further,  
in what they call Landau-Potts perturbative field theory, for arbitrary values of $q$. 
They take a quintic potential which for $q = 3$ is
$$
U \propto ( \phi_1^2 + \phi_2^2 ) \,\phi_2 ( \phi_2^2 - 3\phi_1^2 ).
$$
With the interchange $\phi_1 \leftrightarrow \phi_2$, it is 
just the term $c_4I_1I_2$ of potential (\ref{13pot}). 
The perturbative beta function yields a non-trivial fixed point of the type 
previously found in the single-field case \cite{Blume-Capel}. 
The coupling constant is of order $\sqrt{-\ve}$, as generally occurs 
in the $\ve$-expansion for potentials with leading odd terms \cite{N-C-Stanley}. 

Nicoll, Chang, and Stanley \cite{N-C-Stanley}
study bifurcations from the Gaussian solution of the isotropic 
vector model, in terms of solutions of a linearized ERG (\ref{ERG}). 
The $S_3$ anisotropy complicates the beta functions, but
the nature of the problem is the same. 
For the sake of simplicity, let us now 
consider a truncation at level $M$, that is to say, to a number $M$ of coupling constants. 
Bifurcations of solutions of the beta-function equations occur when the Jacobian matrix
$\p_i\b_j$ 
has a zero eigenvalue. 
If we exclude the first one, the beta-functions have the structure
\begin{equation}
\b_i = \k_i\,(d_i-d)\,c_i + p\!\left[(1+2c_1)^{-1},c_2,\ldots,c_M\right],
\quad i=2,\ldots,M,
\label{bfun}
\end{equation}
where $d_i$ is a fractional number,
$\k_i$ is a positive number, and 
the polynomial $p$ is a sum of monomials of several degrees, including some 
of first degree. These ones always involve $c_j$ with $j>i$. Therefore,  
the Jacobian matrix is upper triangular at the point $c_i=0,\;i=1,\ldots,M$. 
The eigenvalues of an upper triangular matrix are the diagonal terms of the matrix.     
In our case, one of them vanishes whenever $d=d_i$. The series of critical dimensions 
that follows the labeling in Eq.~(\ref{13pot}), is: $6,4,10/3,3,14/5,8/3,\ldots$, and 
continues with more values $d_i < 3$. 
Naturally, this series of critical dimensions is given by 
the degrees of successive powers of fields in the potential, namely, the series  
$d_n=2n/(n-2),\;n=3,4,5,6,\ldots.$ 

We are interested here in the bifurcation at $d_\mathrm{c}=10/3$, corresponding to 
the fifth-degree term ${c}_{4}I_1I_2$ in Eq.~(\ref{13pot}). It is  the only one with 
$d_\mathrm{c} > 3$, apart from $d_\mathrm{c}=6$ or 4. 
As the non-trivial fixed point departs from the Gaussian fixed point for 
growing $\ve=10/3-d$, the fixed-point value ${c}_{4}^*$ also grows (in absolute value). 
So do the first $O(2)$-breaking coupling ${c}_{2}^*$ and the mass coupling ${c}_{1}^*$. 
The behavior of coupling constants near a bifurcation point can be studied with 
the $\ve$-expansion. However, a deeper treatment demands 
standard methods of bifurcation theory 
\cite{Gu-Ho,Go-Sch}, as noticed by Gukov \cite{Gukov}.
Before dealing with the problems that arise for $d_\mathrm{c}=10/3$, let us review the 
ordinary $\ve$-expansion about $d_\mathrm{c}=4$.

\subsection{Expansion in $\ve=4-d$}

Nicoll et al's early study of the $\ve$-expansion of the Wegner-Houghton equation for the $O(N)$ vector model \cite{N-C-Stanley} has been 
revisited by other authors \cite{Comellas-T,Aoki}. 
All these authors are mainly concerned with the calculation of critical exponents, while 
we are also interested in the qualitative properties of the bifurcations.  

In bifurcation theory, the types of bifurcation phenomena are studied by means 
of coordinate transformations that reduce bifurcations to {\em normal forms}, 
that is to say, to standard polynomial functions that are more manageable and, therefore, 
have been extensively studied \cite[\S 3.3]{Gu-Ho}. The simplest normal form is the 
{\em saddle node}, with only one relevant variable $x$ and one bifurcation parameter $\mu$, 
and general expression $\mu - x^2$ 
(up to a sign). 
In this saddle-node bifurcation, 
there are two equilibrium points for positive values of the parameter that 
merge and disappear for negative values.
Any one-parameter bifurcation can be perturbed to become of the saddle-node type, while 
a perturbation of the saddle-node bifurcation leads to another of the same type 
(it is said that saddle-node bifurcation is the only one that is {\em structurally stable}) \cite[\S 3.4]{Gu-Ho}. 
However, this simple normal form can sometimes be ruled out by the setting of the problem, 
as occurs in the present case. 

The Gaussian fixed point, with vanishing coupling constants, 
is present for any value of the dimension (our bifurcation parameter), and this 
condition indeed rules out the saddle-node form. If we relax one of the requirements for 
this form, namely, that the partial derivative with respect to the parameter at 
the bifurcation point is non-vanishing, we obtain the next normal form, called 
{\em transcritical}, with general expression $\mu x - x^2$ \cite[\S 3.4]{Gu-Ho}. 
In this normal form, $x=0$ is always a solution and there is 
another (non-null) solution, $x=\mu$. 
The bifurcation causes a change of stability of the two solutions.

To bring a system of equations to its normal form, 
one has first to make a linear coordinate transformation that diagonalizes 
the Jacobian matrix at the bifurcation point. 
This linear transformation is given by a matrix 
whose rows are left eigenvectors of the Jacobian matrix. Naturally, 
the linear RG equations, once diagonalized, are trivially integrated. 
This procedure was already applied by Wilson and Kogut to their simplified 
renormalization group, which was, in essence, a truncation of the exact 
RG for the $\l\phi^4$ theory to the flow of two parameters, $m^2$ and $\l$, thus involving 
the diagonalization of a $2 \times 2$ matrix 
\cite[\S 4]{Wil-Kog}. 

The linearized Wegner-Houghton equation for the 
$N$-vector model is a partial differential equation 
whose eigenfunctions are the Laguerre polynomials 
$L_n^{N/2-1}$ \cite{N-C-Stanley,Comellas-T}. In particular, 
the eigenfunction corresponding to the eigenvalue $4-d$ is 
$$L_2^{N/2-1}\!\left(\frac{d-2}{2A_d}\, I_1\right)=\frac{1}{8}
        \left[\frac{(d-2)^2}{A_d^2}I_1^2-2(N+2)\frac{d-2}{A_d}I_1+N(N+2)\right],$$ 
being $I_1$ the quadratic field invariant and $A_d = [2^{d-1}\pi^{d/2}\Gamma(d/2)]^{-1}$. 
This eigenfunction involves the first two couplings of the vector model, 
namely, $I_1$ and $I_1^2$. 
Comparing to potential (\ref{13pot}), which includes anisotropic couplings, 
and keeping the same numbering of coupling constants,  
we have the ratio
\begin{equation}
\frac{c_1}{c_3} = \frac{2A_d(N+2)}{2-d}\,.
\label{c1c3}
\end{equation}
For $d=4$, this ratio is $c_1/c_3=-A_4(N+2).$

Regarding beta-functions (\ref{bfun}), 
the simple form of eigenfunctions is a consequence of 
the Jacobian matrix being 
upper-triangular (a property that holds for arbitrary $N$). 
The first two beta-functions are \cite{MOP}: 
\begin{align}
\b_1 &= 2\, c_1 + A_d\frac{2(N+2)}{1+2\, c_1}\,c_3 \,,
\label{b1}\\
\b_3 &= (4-d)\, c_3 - A_d\frac{4(N+8)}{(1+2\, c_1)^2}\,c_3^2\, .
\label{b3}
\end{align}
These beta-functions are equivalent to Wilson and Kogut's simplified renormalization group  
\cite[\S 4]{Wil-Kog} and 
suffice to calculate the Wilson-Fisher fixed point 
to ${\Mfunction O}(\ve)$. It is given by
$$c_3 = \frac{\ve}{4A_4(N+8)},\quad c_1 = -\frac{(N+2)\ve}{4(N+8)}\,.
$$
Hence, $c_1/c_3=-A_4(N+2),$ as above. 
In addition, we can observe that (\ref{b3}) has the transcritical bifurcation normal form 
(up to a numerical coefficient). In fact, the left eigenfunction for 
eigenvalue $4-d$ has only one non-vanishing component, along $c_3$ \cite[\S 4]{Wil-Kog}. 

The first order in $\ve$ can give valuable information about the bifurcation, but 
it may not be reliable for $\ve=1$, of course. Let us check on the reliability of 
the second-order results given by Aoki et al \cite{Aoki}, namely,
\begin{align}
2\, c_1
&= -\frac{(N+2)\,\ve}{2 (N+8)} 
-\frac{(N+2) \left(N^2+50 N+192\right) \ve^2 }{4 (N+8)^3}\,,
\\
\nu
&= \frac{1}{2} + \frac{(N+2)\,\ve}{4 (N+8)} +
\frac{(N+2) \left(N^2+38 N+96\right) \ve^2}{8 (N+8)^3}\,
\end{align}
($c_3$
is omitted here). 
With $\ve=1$ in these formulas, we obtain $2c_1=-0.417$ and $\nu=0.653$ for $N=1$, 
while, for $N=2$, we obtain $2c_1=-0.496$ and $\nu=0.688$.
The exact results of the Wegner-Houghton equation (\ref{ERG}) 
are $2c_1=-0.462$ and $\nu=0.690$, for $N=1$ \cite{Morris_2}, and
$2c_1=-0.537$ and $\nu=0.768$, for $N=2$ \cite{Alford,Comellas-T}.

In conclusion, we have in this case an easily identifiable type of bifurcation and 
a semi-quantitative accuracy of the $\ve$-expansion.

\subsection{Expansion in $\ve=10/3-d$}
\label{f_e-expansion}

The first bifurcation to a 
non-trivial field theory at a non-integer dimension 
is at $d_\mathrm{c}=10/3$. Some results of the expansion about it  
have been obtained in perturbation theory by Codello et al
\cite{Blume-Capel,mcrit-LP} and by Gracey \cite{Gracey}.  
Before examining the expansion in the bifurcation parameter, let us analyze 
what type of bifurcation we now have. 

We have seen that the existence of a solution with null coupling constants, 
in any dimension, forces the partial derivative with respect to 
the dimension to be null at the bifurcation point, 
giving rise to the transcritical bifurcation. 
However, there can be further natural constraints in the equations, such 
as the presence of a symmetry in them \cite[\S 3.4]{Gu-Ho}. 
The simplest symmetry is a reflection symmetry, which we indeed have in the Potts model 
beta functions. 
This symmetry is due to the invariance of potential (\ref{13pot}) 
[or extensions thereof, e.g.~(\ref{18pot})]
under the change of sign of field $\phi_1$, 
{provided that all the signs of $c_2$, $c_4$, $c_7$, etc, are also reversed.}
Therefore, the beta functions and the fixed-point equations 
are invariant under changes of sign of $c_2$, $c_4$, etc. The equation 
that contains the vanishing eigenvalue at $d_\mathrm{c}=10/3$ is $\b_4=0$, and $\b_4$ is 
odd under the symmetry, which implies that its second derivative with respect to 
$c_4$ cannot vanish. This is a new constraint that rules out the transcritical bifurcation 
and leads to the next case, namely, the {\em pitchfork bifurcation} 
\cite[\S 3.4]{Gu-Ho} or \cite[\S I.1]{Go-Sch}.

The normal form of the pitchfork bifurcation is $\mu x - x^3$ \cite[\S 3.4]{Gu-Ho}. 
Unlike the transcritical bifurcation, this bifurcation is not common 
in the RG literature, although it is discussed by Gukov \cite{Gukov}.
At the bifurcation point of the pitchfork bifurcation, the stability 
of the trivial solution $x=0$ changes, as there appear a new pair of stable solutions, 
related by the reflection symmetry, namely, $x=\pm\sqrt{\mu}$. The graph of the bifurcation 
in the $(\mu,x)$ plane resembles a pitchfork. In our case, 
the two stable symmetry-related solutions, 
with non-vanishing values of $c_2$ and $c_4$, correspond to 
$O(2)$-breaking but $S_3$-invariant RG fixed points.

It is possible to calculate the linear eigenfunctions, as is done in the $\ve$-expansion 
of the vector model at $d_\mathrm{c}=4$. They are still Laguerre polynomials, 
but multiplied now by harmonic polynomials to account for the anisotropy. 
Anisotropic solutions of the linearized approximate recursion formula
for the vector model were already 
introduced by Wegner in 1972 \cite{harmonic}. Those solutions can be employed here. 
For $N=2$, the eigenfunction corresponding to the eigenvalue $5-3d/2$ is 
\begin{equation}
L_1^3\!\left(\frac{d-2}{2A_d}\, I_1\right) P_3(\phi_1,\phi_2) = 
\left(4-\frac{d-2}{2A_d}\, I_1\right) I_2\,,
\label{eigenf}
\end{equation}
where $P_3(\phi_1,\phi_2)=\phi_{1}^{3}-3\phi_{1}\phi_{2}^{2}=I_2$ 
is the third-degree harmonic polynomial. Comparing expression (\ref{eigenf}) to 
the corresponding part of potential (\ref{13pot}), namely, 
${c}_{2}I_2+{c}_{4}I_1I_2$, we obtain the ratio 
\begin{equation}
\frac{c_2}{c_4} = \frac{8A_d}{2-d}\,.
\label{c2c4}
\end{equation}
For $d_\mathrm{c}=10/3$,
$${c}_{2}/{c}_{4}=-6A_{10/3}\,.$$  

Given that the equation that contains the vanishing eigenvalue at $d_\mathrm{c}=10/3$ is 
$\b_4=0$, the minimal truncation involves $\b_i(c_1,\ldots,c_4),\; i=1,\ldots, 4.$ 
Let us assume, for the moment being, that there are solutions such that $c_1=c_3=0$,  
so that we only have to solve 
for $c_2$ and $c_4$ the following equations: 
\begin{align}
\b_2 &= \left(3-\frac{d}{2}\right) c_2 + A_d\, 8\,c_4 = 0\,,
\label{b2}\\
\b_4 &= \left(5-\frac{3d}{2}\right) c_4 + A_d\, 288\,c_2^2c_4 = 0\,,
\label{b4}
\end{align}
They yield, at the lowest order in $\ve$,
$$
c_2 = \frac{\sqrt{-\ve}}{8\sqrt{3}\,A_{10/3}^{1/2}}, \quad 
c_4 = \frac{-\sqrt{-\ve}}{48 \sqrt{3}\,A_{10/3}^{3/2}},
$$
or the solution with interchanged signs of $c_2$ and $c_4$ (which is physically equivalent).
These two solutions reproduce the above-quoted ratio ${c}_{2}/{c}_{4}=-6A_{10/3}.$
In addition, 
we can substitute ${c}_{2}=-6A_{10/3}{c}_{4}$
in $\b_4$ 
to get
\begin{equation}
\b_4 = \frac{3\,\ve}{2}\, c_4 + A_{10/3}^3 10368\,c_4^3,
\label{bpitch4}
\end{equation}
having the pitchfork bifurcation normal form (up to numerical coefficients). 
Let us remark that perturbation theory also leads to the pitchfork form 
[see Eq.~(\ref{betaR}) in the appendix, 
with an outline of the calculation].

However, 
there is actually no {(nontrivial)} solution such that $c_1=c_3=0$, and the non-null values 
of $c_1$ and $c_3$ contribute to 
the lowest order in $\ve$, in spite of the fact that they are themselves of higher order 
than $c_2$ and $c_4$. They contribute, for example, through the 
quadratic term $c_3\,c_4$ that is present in $\b_4$. 
Thus, the lowest order in $\ve$ involves, at the least, all the 
equations $\b_i(c_1,\ldots,c_4)=0,\; i=1,\ldots, 4.$ The coupling 
constants $c_i,\, i=1,\ldots,4$, are precisely the RG-relevant ones at the trivial 
fixed point for $d<10/3$. 
Since the fixed-point equations can be expressed as polynomial equations, 
it is not difficult to see that we have a solution 
such that $c_1$ and $c_3$ are ${\Mfunction O}(\ve)$ and 
$c_2$ and $c_4$ are ${\Mfunction O}(\sqrt{\ve})$. Actually, after some calculations, 
we obtain
\begin{equation}
c_1 = \frac{-27\,\ve}{512} , \quad 
c_2 = \frac{\sqrt{-\ve}}{16\sqrt{6}\,A_{10/3}^{1/2}}, \quad 
c_3 = \frac{3\,\ve}{256\,A_{10/3}}, \quad
c_4 = \frac{-\sqrt{-\ve}}{96 \sqrt{6}\,A_{10/3}^{3/2}}
\label{e4}
\end{equation}
(or the solution with interchanged signs of $c_2$ and $c_4$). 

Further calculations 
indicate that this lowest order solution is modified in higher-order 
truncations. In fact, the signs of $c_1$ and $c_3$ turn out to be wrong, while 
$c_2$ and $c_4$ are less modified. 
Of course, 
${c}_{2}/{c}_{4}=-6A_{10/3}$ is always fulfilled. 
The modification is due to the non-negligible effect of the new coupling constants, 
because  
this $\ve$ expansion is not as simple as the one for $\ve=4-d$ in the $O(N)$ model. 
Now, the left eigenvector for the eigenvalue $5-3d/2$ has several components,  
one along $c_7$.  
To wit, the variable associated with the diagonalization of the Jacobian matrix is 
\begin{equation}
c'_4 = 
c_4 + \frac{20}{d-2}\,c_7 + \cdots 
\label{c4p}
\end{equation}
Therefore, Eq.~(\ref{bpitch4}), 
or the improved version that takes $c_1$ and $c_3$ into account, 
need to be substituted by another equation with $c_4$ replaced by $c'_4$. 
In fact, we have to enlarge the set of equations so as 
to include $c_7$ to obtain the right signs of $c_1$ and $c_3$ in the 
$\ve$ expansion.

The above complications may question our asserting that the bifurcation is the pitchfork 
rather than a more complicated one with the same symmetry. 
To establish a normal form with mathematical rigor, 
it is necessary to have a {\em transversality} condition. 
For the pitchfork, it is necessary to have 
a non-vanishing third derivative with respect to the relevant variable. 
The relevant variable is the one that we have denoted $c'_4$, to  be obtained from 
the full expression of the left eigenvector. To work out the normal form,  
we could also follow the general procedure of {\em Lyapunov-Schmidt reduction} 
\cite[\S I.3]{Go-Sch}. To wit, we could leave $c_4$ as such and consider 
a new function $\b_4(c_4)$ where the 
couplings constants $c_i=0,\;i\neq 4,$ are expressed in terms of $c_4$ by 
solving the equations $\b_i=0,\;i\neq 4$. 
However, a proof of the transversality condition that is independent of the truncation
demands us to work with the partial differential equation and  
calculate  
its full third functional derivative.  
This calculation is by no means straightforward, and therefore it is difficult to have 
a mathematically rigorous proof of the type of bifurcation in our case. 

Moreover, assuming that we have a pitchfork bifurcation  
and a lowest order such that $c_1$ and $c_3$ are ${\Mfunction O}(\ve)$ and
$c_2$ and $c_4$ are ${\Mfunction O}(\sqrt{-\ve})$, respectively, 
we still have an obvious problem: 
the $\ve$-expansion gives imaginary values of $c_2$ and $c_4$ 
for $\ve>0$ and, therefore, it is only applicable to $d>10/3$. 
This problem forbids its use in $d=3$. 
We now turn to a different kind of analysis. 

\section{Analysis of the RG fixed point with origin at $d_\mathrm{c}=10/3$}
\label{S3FP}

Going beyond the $\ve$-expansion does not seem {feasible}, unless 
we employ a truncated potential. 
Thus, let us examine, in a finite-dimensional space of couplings, 
both the type of bifurcation at $d_\mathrm{c}=10/3$ and, especially, the dependence of 
the non-trivial fixed points with $d$. 
By letting $d$ vary, we are dealing with one variable more.  
Employing geometrical concepts, as in Sect.~\ref{RG}, 
the system of algebraic equations $\b_i=0,\;i=1,\ldots,M,$ 
defines a set of algebraic surfaces in the $M+1$-dimensional space, whose intersection 
is a curve that consists of the solutions of the system of equations 
parametrized by $d$. This type of curve is called, in algebraic geometry, 
a unidimensional {\em complete intersection variety}, and it    
has several branches, because of the bifurcations at the critical dimensions
(apart from spurious parts). 
To wit, the algebraic curve has the trivial branch  
$c_i=0,\;i=1,\ldots,M$ (a straight line), with singular points 
at the critical dimensions, where bifurcations occur. We intend 
to examine the non-trivial branch that arises at $d_\mathrm{c}=10/3$.

To simplify somewhat the Wegner-Houghton equations and hence the beta-functions, 
we make the usual redefinition of the potential as $U \ra A_d\,U$ 
and the fields as $\phi_i \ra A_d^{1/2}\phi_i$, 
with $A_d = [2^{d-1}\pi^{d/2}\Gamma(d/2)]^{-1}$. 
Thus, this irrational number disappears from the beta-functions 
(as by making $A_d\ra 1$),    
giving rise to rational polynomial equations for the fixed points. 
Nevertheless, the polynomials are long and complicated even for low values of $M$.

\subsection{Truncation of the potential at $\phi^{8}$}
\label{phi8}

At first, we truncate the potential at $\phi^{8}$, 
that is to say, we take the part of potential (\ref{13pot}) that includes up to $c_9$
(spanning, with $d$, a ten-dimensional space). Notice that we thus include 
two field powers that are RG-irrelevant at the trivial fixed point in $d=3$, 
namely, $\phi^{7}$ and $\phi^{8}$. It turns out that 
this potential is sufficient to have an accurate ${\ve}$ expansion at $d_\mathrm{c}=10/3$ 
and, in addition, allows us to find the most important feature of 
the non-trivial solution branch. 
We have tried truncations at lower powers of the field and found that they are 
insufficient for these purposes. 

\begin{figure}
\includegraphics[width=8cm]{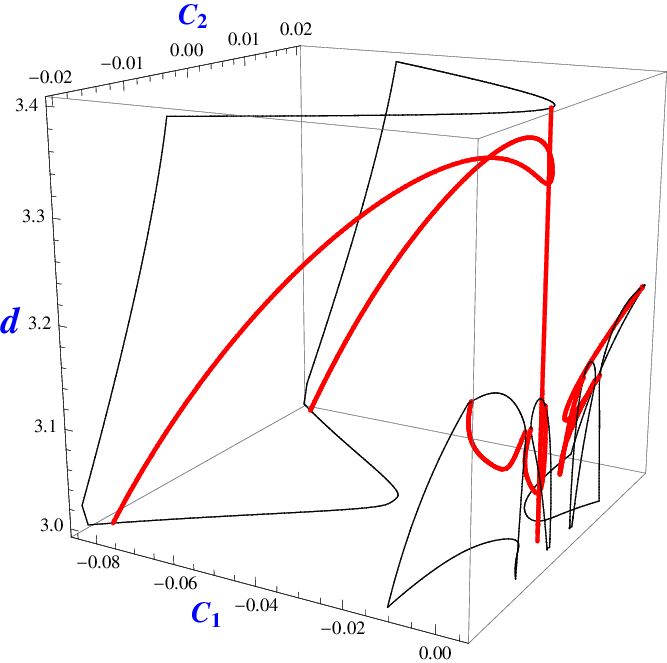}
\caption{Pitchfork bifurcation at $d_\mathrm{c}=10/3$ in a three-dimensional projection 
of the RG fixed-point curve
corresponding to truncation at nine coupling constants.}
\label{pfork}
\end{figure}

The nine algebraic equations 
derived from $\b_i=0,\;i=1,\ldots,9,$ occupy more than 56 lines and constitute 
a formidable system. 
We have been able to diminish its complexity by eliminating the higher coupling constants 
in favor of the lower ones, thus reducing it to a system of two equations in $c_1$, $c_2$ 
and $d$, which occupies some more space, namely, 76 lines, but is actually easier 
to handle. After this dimensional reduction, 
we now have, geometrically speaking, a complete intersection of 
two algebraic surfaces (of high degree) in a three-dimensional space, 
giving rise to the curve that we seek (that is, giving rise to 
its projection onto the three-dimensional space). In Fig.~\ref{pfork}, 
we show the plot of the branches of the curve in relevant intervals of the variables 
(plus some spurious part of the curve). 

Although the bifurcating equation is $\b_4=0$, with a vanishing eigenvalue 
affecting $c_4$, we can replace $c_4$ by $c_2$ to display de pitchfork bifurcation, 
in accord with their relationship at the bifurcation point, namely, 
$c_2=-6c_4$ (Sect.~\ref{f_e-expansion}).
It is apparent in Fig.~\ref{pfork} that 
the normal form is actually $\ve c_4 + c_4^3$ instead of $\ve c_4 - c_4^3$, 
having three solutions in the neighborhood of the bifurcation point 
for $d>10/3$ rather than for $d<10/3$.  
This agrees with 
the results of Sect.~\ref{f_e-expansion}. 
However, a very remarkable property of the non-trivial branch is that 
it reaches a maximum $d$, at some dimension $d>10/3$, and 
then turns to decreasing $d$, eventually coming down to $d=3$. The turning point is 
a saddle-node bifurcation point. It can be found, in principle, 
by solving the two equations in $c_1$ and $c_2$ plus the condition of vanishing 
(two-dimensional) Jacobian. However, the calculation is very complicated, 
due to the high degree of the two polynomial equations.

Naturally, the normal form of a bifurcation is only valid in a neighborhood of 
the bifurcation point and 
the behavior of the curve far from that point is independent of the normal form.  
Certainly, the form of the pitchfork bifurcation at $d_\mathrm{c}=10/3$ 
does not tell us what happens at $d=3$. 
We see that the $\ve$-expansion is misleading in this case, because an expansion 
in powers of $\sqrt{-\ve}$ is only valid for $\ve<0$ (as a real function). 
Thus, the saddle-node bifurcation point must occur as a 
singularity of the $\ve$-expansion.  
In contrast,    
the $\ve$-expansion at $d_\mathrm{c}=4$ for the $O(N)$ vector model
does not have such type of singularities, since the curve corresponding to the 
Wilson-Fisher RG fixed point goes down to $d=3$ monotonically (although the $\ve$-expansion  is only asymptotic, of course). 
 
Apart from the saddle-node bifurcation, there are no more bifurcation points as we proceed along the curve, down to $d = 3$ 
(Fig.~\ref{pfork}).
The location and type of the bifurcation points tell us where and how the stability of 
the RG fixed points changes with $d$. 
Let us write the beta function
as
$$\dot{c}=\ve c + c^3,$$ 
where $c$ is to be identified, up to a numerical factor, with either $c_4$ or $c_2$ 
(which are linearly related, close to the bifurcation point).
The positive sign determines that it  
has the so-called {\em subcritical} pitchfork form 
rather than the {\em supercritical} form \cite[\S 3.4]{Gu-Ho}. 
The former is such that there are two unstable RG fixed points and one stable for $\ve<0$
while the only fixed point for $\ve>0$ is unstable. In our case, this is confirmed 
by computing (numerically) the Jacobian matrix eigenvalues, which include,  
for $d$ larger than but close to $10/3$,  
four positive eigenvalues, one more than the stable solution. 
Of course, the null branch of the Gaussian fixed point always 
has three positive eigenvalues for $d>10/3$ and four for $d<10/3$, as corresponds 
to naive power counting. 

Naturally, the stability of fixed points on the non-trivial branch 
changes again at the saddle-node bifurcation point, namely, 
at the new critical dimension that is present for some 
$d_\mathrm{c}>10/3$ (Fig.~\ref{pfork}). There, the unstable fixed point becomes stable again, as 
the Jacobian matrix loses one positive eigenvalue, a fact that we have confirmed 
numerically.

No further bifurcations occur on the non-trivial branch. Therefore, we have, for $d<d_\mathrm{c}=10/3$, a four-fold unstable Gaussian fixed point and two additional fixed points with one less instability, which are placed symmetrically an are physically equivalent.  Presumably, the Gaussian fixed point is joined by a RG trajectory to each of the two non-trivial fixed point.

In regard to the RG flow resulting from this combination of bifurcations, let us recall that 
Gukov \cite{Gukov} considers the possibility of 
a subcritical pitchfork bifurcation {\em stabilized} by means of fifth-order terms, namely, 
one with form 
$$\ve c+ c^3-c^5$$ 
($c$ being a generic coupling, in our case, $c_4$ or $c_2$).
This form is not really a local normal form 
different from $\ve c+ c^3$ at $\ve=c=0$, 
although it is a useful expression that encompasses both the pitchfork and 
the two saddle-node bifurcations. 
In fact, $\b_4$ may adopt this form after a Lyapunov-Schmidt reduction, namely, after 
solving the equations $\b_i=0,\;i\neq 4,$ for $c_i,\;i\neq 4,$ substituting in $\b_4$, 
expanding in powers of $\ve$ and $c_4$, and neglecting higher-order terms. 

If we assume that the bifurcation equation, 
truncated at ${\Mfunction O}(c^5)$, has indeed Gukov's form and, 
to be definite, is
$$
\b(c,\ve) = \ve c+ c^3-\a c^5=0,\;\;\a>0,
$$
then we can obtain the non-trivial branch and its $\ve$-expansion. Factoring $c$ out, solving the biquadratic equation, and then expanding in $\ve$, we have
$$c = \sqrt{\frac{1}{2\a}\left(1-\sqrt{1 + 4\a \ve}\right)} =
\sqrt{-\ve} \left(1- \frac{\a \ve}{2} + {\Mfunction O}(\ve^2) \right),
$$
where we have selected the positive solution that vanishes at $\ve=0$. 
We can see that the function of $c(\ve)$ 
is real only for $\ve<0$ and, besides, has a singular point at $\ve=-1/(4\a)$, 
where $dc/d\ve \ra \infty$.
This singularity corresponds to the saddle-node bifurcation and does not appear in 
the $\ve$-expansion.

Since we have found that the new non-trivial fixed point has three unstable directions, 
it is less stable than the Wilson-Fisher fixed point, with only two relevant directions. 
Thus, the fully symmetric behavior is favored. 
At any rate, this $S_3$-invariant fixed point in $d=3$ 
is not a spurious fixed point (in the sense of Margaritis et al \cite{MOP})
but a truly real fixed point. 
This fact is further demonstrated by examining higher order truncations.

\subsection{Tracing the fixed point with up to 18 coupling constants}
\label{18}

As already noticed, the beta functions of higher order couplings give rise to long 
high-degree polynomial equations, which are increasingly 
complex. With more couplings, 
we expect that the pitchfork structure is preserved near the bifurcation point, 
but the non-trivial fixed-point branch is likely to change far from the bifurcation point. 
In any truncation of the coupling-constant space, the non-trivial branch of the curve is 
a complete intersection variety. 
However, the algebraic elimination of variables soon becomes unfeasible (with current 
algebraic computing systems). 
We need to resort to non-algebraic methods, mostly based on numerical computation, 
whose accuracy must be monitored. We have found that the method of 
intersection tracing is especially useful. We now describe this method briefly 
(for a detailed description, see \cite{tracing}).

\begin{figure}
\includegraphics[width=8cm]{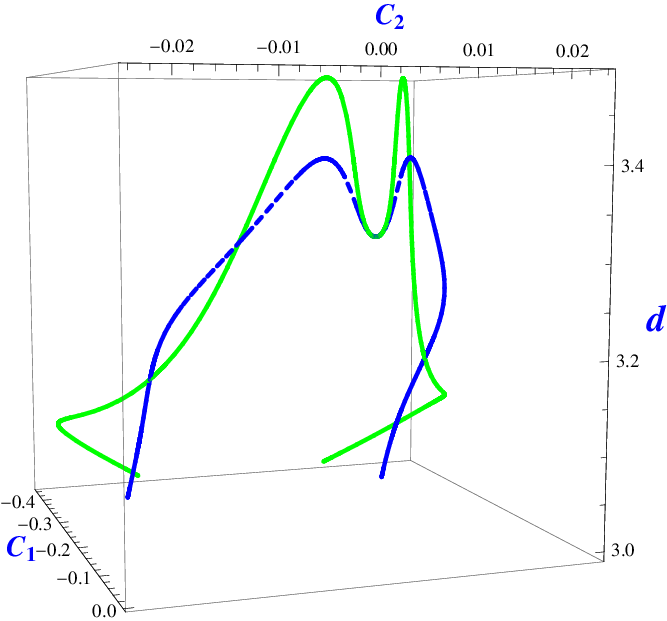}
\caption{The non-trivial branch 
of the pitchfork bifurcation at $d_\mathrm{c}=10/3$, 
computed with truncations at $13$ or $18$ coupling constants (dashed blue line or solid green line, respectively).}
\label{pfork1}
\end{figure}

Intersection tracing consists in constructing, starting from a known point of a 
complete-intersection curve, 
a local approximation of the curve given by the curve tangent. 
By stepping along the approximation a specific distance, we obtain
an estimate of the next curve point, where we can again compute the tangent. 
Performing this procedure a sufficient number of times, we follow the chosen 
branch of the curve up to the desired point, at $d=3$. 
We monitor the accuracy of the procedure 
by computing the values of $\b_i,\;i=1,\ldots,M,$ at successive points, 
and checking that their absolute values stay below an adequate bound. 

We start the tracing at the bifurcation point.
At this point, the curve has a double tangent, and we have to choose the one  
corresponding to the non-trivial branch. 
It is given by eigenfunction (\ref{eigenf}), with non-null 
${c}_{2}$ and ${c}_{4}$, and the ratio ${c}_{2}/{c}_{4}=-6$ (now $A_d\ra 1$).
The results of the tracing for the truncations with $M=13$ or 18 coupling constants 
are displayed in Fig.~\ref{pfork1}, projected onto the same three-dimensional space as 
in Sect.~\ref{phi8}. 

The above procedure is remarkably successful. In general, 
it allows us to avoid the multitude of spurious solutions of the algebraic equations 
at low $d$ ($d\geq 3$) and 
isolate the real and interesting ones. Of course, we can only guarantee that 
the solutions are accurate at a definite truncation level, as they change somewhat 
from one level to another, except when examined very close to the bifurcation point 
(Fig.~\ref{pfork1}). 
Presumably, they converge as the truncation level increases. 
The full procedure is a mixture of algebraic and numeric computations, as it
begins with the algebraic calculation of beta functions.  
This algebraic calculation does not take long for 18 couplings, 
although the 18-coupling beta functions occupy about 
{$300$ text lines (that is about five pages, the exact value depending on the format).} 
Unfortunately, the algebraic calculation becomes very cumbersome and hardly feasible
for higher-level truncations.

Let us show some characteristics of the $S_3$ fixed point with 18 couplings, at $d=3$
(from the two physically equivalent points, let us take the one with $c_2^*>0$).  
The vector of coupling constants is: 
\begin{align*}
(-0.221,0.0123,0.0314,0.000792,-0.0000473,0.00282,-0.000284,-0.
   0000227,0.0000412,\\-1.32\cdot 10^{-7}, -0.0000686, -1.83\cdot 10^{-6}, -0.0000181, 
   4.77\cdot 10^{-8}, 
-4.09\cdot 10^{-6}, 1.20\cdot 10^{-9},\\ 2.77\cdot 10^{-7}, -9.74\cdot 10^{-7}).
\end{align*}
The precision is reduced for clarity. Using the full precision values, the beta-functions 
vanish to machine-precision.
The first six eigenvalues (dimensions) are:
$$\{1.63, 0.622, 0.501, -0.726,-1.00,-1.41\}.$$
The other eigenvalues are more irrelevant than the ones displayed. 
The eigenvectors that correspond to the relevant (positive) eigenvalues, in descending order, are:
\begin{align*}
(0.989, 0.112, 0.0928, 0.0142, 0.000219, 0.00580, 0.000613, 0.0000408, \ldots),\\
(0.438, -0.889, -0.0825, -0.107, 0.00152, -0.00675, 0.00737, 0.00151, \ldots),\\
(0.124, -0.982, -0.0967, -0.106, 0.00274, -0.00202, 0.0118, 0.00202, \ldots).
\end{align*}
The most relevant eigenvector basically goes along the ``mass direction'', namely, 
along the vector $(1,0,0,\ldots)$. In contrast, the third eigenvector basically
goes along the ``$S_3$ direction''. As the fixed point is such that $c_1^*<0$, 
we have oriented the eigenvectors in the direction of increasing $c_1$. Given that 
$c_2^*>0$, $c_2$ decreases in the direction of the third eigenvector, and 
the same happens to $c_4$.

The above quoted eigenvalues and eigenvectors correspond to 
the current truncation level and should change somewhat with the truncation level. However, 
let us emphasize that the number of relevant eigenvalues is associated with the type of bifurcation and is 
therefore preserved.

{\section{
The bifurcation at $d_\mathrm{c}=10/3$ with other regulators}
}
\label{Litim}

The preceding analysis is based on the sharp-cutoff Wegner-Houghton ERG, but 
it is interesting to know if the bifurcation pattern at $d_\mathrm{c}=10/3$ is universal, 
namely, if it is independent of the regularization scheme. 
We focus on an alternative regulator that is employed in recent analyses 
\cite{BAZ,S-V}. 

{
With this regulator (often named after Litim in the literature), 
the equation for the effective potential that replaces Eq.~(\ref{ERG}) is
\begin{equation}
\frac{\p U(\phi_\a,\L)}{\p \L} = B_d \,\L^{d+1}\,
\mathrm{tr}\left[\L^2 \d_{\a\b}+ U_{\a\b}(\phi_\a,\L)\right]^{-1},
\label{LERG}
\end{equation}
where $\a,\b=1,\ldots,N$ and $B_d = [2^{d-1}\pi^{d/2}d\,\Gamma(d/2)]^{-1} =  A_d/d$.
With a redefinition of the potential as $U \ra B_d\,U$ 
and the fields as $\phi_i \ra B_d^{1/2}\phi_i$, the constant $B_d$ disappears and, for $N=2$, 
equation (\ref{LERG}) becomes \cite{BAZ}:
\begin{equation}
\frac{\p U(\phi_\a,\L)}{\p \L} = \L^{d+1}\,
\frac{2\L^2 + \mathrm{tr}\,U}{\L^4 + \L^2 \,\mathrm{tr}\,U + \det U}\,.
\end{equation}
We can employ the expansion of the potential $U$ in powers of $I_1$ and $I_2$
and obtain a set of coupling constants, as in Eq.~(\ref{13pot}). 
However, the beta-functions $\b_i$ have now different expressions. They 
still conform to the structure (\ref{bfun}), but the polinomials $p$ are different.
In the appendix, sect.~\ref{betas}, a few initial beta-functions with either regulator 
are displayed, for comparison. It is evident that the new beta-functions are 
more complex, because the power of $(1+2c_1)^{-1}$ in every polinomial $p$ is of one 
degree higher. This added complexity further restricts the maximum truncation level
that is feasible.
}

\begin{figure}
\includegraphics[width=8cm]{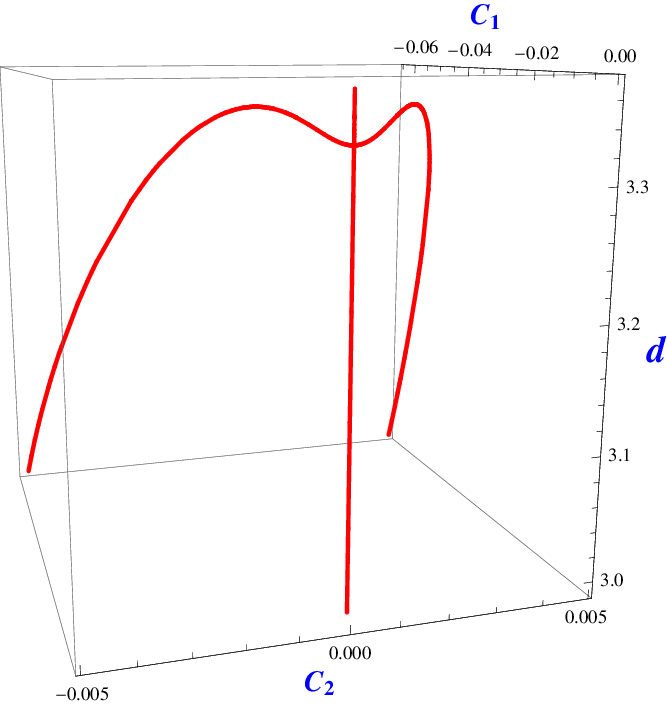}
{\caption{Pitchfork bifurcation computed with another regulator 
and nine coupling constants.}
}
\label{pforkL}
\end{figure}

{
At any rate, our purpose here is just to confirm that the bifurcation pattern 
already found with the sharp-cutoff regulator is universal. We can do this with a 
limited number of coupling constants, say, with nine of them. To do it, we employ the 
efficient procedure of intersection tracing, starting from the bifurcation point. 
The eigenfunction for the non-trivial branch is now given by a different  
ratio ${c}_{2}/{c}_{4}$, but, apart from this, the procedure is totally analogous. 
The result of the tracing for the nine coupling constants 
is displayed in Fig.~\ref{pforkL}, projected onto the same three-dimensional space as 
before.
} 

{
It is evident that the bifurcation pattern in Fig.~\ref{pforkL} is very analogous to the one 
in Fig.~\ref{pfork}, which has the same namber of coupling constants. 
The present fixed-point  values of the coupling constants at $d=3$ are different, of course.  Actually, they are different even close to the bifurcation point, because the respective 
eigenfunctions are different. This difference is natural, given that fixed-point  
values of coupling constants as well as the eigenfunctions are not universal and are expected to depend on the regulator.
In contrast, the eigenvalues of the Jacobian matrix, namely, the dimensions of 
composite fields must be universal. Therefore, we could compare the respective sets of 
dimensions. However, a comparison with only nine coupling constants is not meaningful, 
because the dimensions at this truncation level are very inaccurate. A comparison with 18  
coupling constants should be more reliable, although probably not very much. At any rate, 
the beta-functions for 18 coupling constants are too complicated in 
the present regularization scheme.
}

{
The preceding discussion should prompt us to seek a better scheme. Surely, there are better 
regulators as regards convergence of truncations. Nevertheless, most 
regularization schemes lead to ERG equations that have, 
unlike Eqs.~(\ref{ERG}) or (\ref{LERG}), an additional integration over 
the regulating function. This integration is 
transferred to the beta-functions, and this complication surely outweighs 
a possible faster convergence of truncations. At any rate, a detailed comparison of 
regularization schemes is beyond the scope of the present study. The next setions 
refer to the sharp-cutoff solutions.
}

\section{Significance of the new RG fixed point}
\label{RG-FP}

Let us consider the significance of the new RG fixed point for 
general statistical models with $S_3$ symmetry. 
We would like to know how it relates to the trivial fixed point and to 
the Wilson-Fisher fixed point. However, we have to keep in mind that 
the relationship between fixed points and the RG flow linking them 
can be easily established only 
in the neighborhood of a bifurcation point. In our case, in $d=3$, the fixed points are fully separated. 
Nevertheless, one can resort to techniques of global analysis 
\cite{N-C-Stanley_1}. For example, the concept of {\em separatrix} is useful. 

In a two-dimensional dynamical system or flow, a separatrix is a curve
that connects the unstable and stable manifolds of a pair of hyperbolic fixed points
(manifolds that are also curves). It defines the boundary between flow trajectories with different qualitative properties. Of course, this 
definition can somehow be generalized to higher-dimensional flows. Indeed, Nicoll et al 
\cite{N-C-Stanley_1} speak of a ``separ-surface'' in their three-dimensional example.
What is called the critical manifold of a 
RG fixed point, in standard RG terminology, is actually a separatrix, 
because all the trajectories leading to the RG fixed point 
form its stable manifold. 
Naturally, a perturbation of the initial condition off the critical manifold give 
rise to non-zero mass and qualitatively different RG trajectories.

Since we have, in our case, that the minimum number of relevant (unstable) directions is two, 
as corresponds to the $O(2)$-invariant RG fixed point, the critical manifold has
codimension two in a neighborhood of that point.  
In contrast, the $S_3$-invariant RG fixed point has a critical manifold with 
codimension three. In the 18-dimensional space of section \ref{18}, 
the tangent space to this manifold is spanned by the negative-eigenvalue eigenvectors 
(the ones not displayed). 
For ease of visualization, let us think of 
the four-dimensional space spanned by 
$c_i,\;i=1,\ldots,4,$ that is to say, the space of relevant coupling constants at 
the Gaussian fixed point in $d=3$. In this four-dimensional space, 
the codimension-two critical manifold at the $O(2)$-invariant RG fixed point 
is a surface, while the codimension-three critical manifold at the 
$S_3$-invariant RG fixed point is a line. 
Presumably, this line is contained in the surface and links both fixed points. 
What we actually know are the tangent spaces at the fixed points only.
Of course, we would like to piece together the results of the local analysis 
around each RG fixed point \cite{N-C-Stanley_1}.

\subsection{Critical manifold}
\label{critman}

Given a set of initial coupling constants, the critical manifold can be found, 
in principle, by trial and error, that is, by tuning the relevant parameters and 
numerically integrating the RG over $t$, 
until a fixed point is reached (for large $t$). This procedure was successfully employed by 
Hasenfratz and Hasenfratz \cite{Hasen2} for the theory of a single scalar field, 
with only one relevant parameter and only the Wilson-Fisher fixed point (they were able to compute a critical value of $c_1$ with seven-digit precision). 
If there are several fixed points and the most stable one has just one relevant parameter, 
that method can still be successful and at least allow us to gather some information 
about the critical manifold. 
However, we know that the most stable fixed point has two relevant parameters and 
we are actually interested in another fixed point that is even less stable.
Thus, tuning only one parameter is ineffective, as attested by Golner's failed attempt \cite{Golner73}.

To gather information about the critical manifold, we can first consider 
the neighborhood of the trivial fixed point at the origin (in $d=3$). 
As has been shown for the single scalar field theory, 
the ERG lends itself to a simpler analysis and a comparison 
with perturbative renormalization in that region \cite{II}. 
In particular, one finds that the linearized ERG provides useful information, 
in particular, an approximation of the critical manifold \cite{II}. 
In the present case, 
the linear ERG equations are given by the Jacobian matrix for null coupling constants. 
This matrix is diagonalized by means of the left-eigenvector matrix transformation. 
Thus, the integration of the linear ERG equations, restricted to 
the space of relevant coupling constants at the origin, yields three simple relations, 
namely,
\begin{align}
c_1 + 8\, c_3 &= K\,c_3^2\,,
\label{RGline1}\\
c_2 + 8\, c_4 &= Q\,c_4^3\,,
\label{RGline2}\\
c_4 &= R\,c_3^{1/2}\,,
\label{RGline3}
\end{align}
where $K$, $Q$ and $R$ are integration constants. 
These three equations are the implicit equations of a trajectory
of the linear ERG, determined by the set of constants $\{K,Q,R\}$. 
The left-hand sides of Eqs.~(\ref{RGline1}) and (\ref{RGline2})
are related to Eqs.~(\ref{c1c3}) and (\ref{c2c4}), with $N=2$ and $d=3$, 
because the latter equations define the right-eigenvectors
(now normalized as if $A_d=1$).

The initial evolution of the coupling constants from some small values is determined by 
the linear ERG and Eqs.~(\ref{RGline1}), (\ref{RGline2}), and (\ref{RGline3}). 
Actually, the linear ERG makes $c_3$ and $c_4$ evolve trivially, in accord with 
their respective dimensions (at the trivial fixed point in $d=3$). This evolution is compatible with 
Eq.~(\ref{RGline3}),  while the evolution of 
$c_1$ and $c_2$ is deduced from Eqs.~(\ref{RGline1}) and (\ref{RGline2}). 
In addition, it is useful to obtain   
the evolution of dimensionful coupling constants, namely, 
$C_n(\L)=\L^{(5-n)/2}\,c_n(t),\;n=1,\ldots,4$. Clearly, 
$C_3$ and $C_4$ do not change with $\L$, while, 
from equations (\ref{RGline1}) and (\ref{RGline2}), we obtain:
\begin{align}
C_1(\L) = C_1(\L_0)-8\,(\L-\L_0)\,C_3,
\label{RGevol1}\\
C_2(\L) = C_2(\L_0)-8\,(\L-\L_0)\,C_4.
\label{RGevol2}
\end{align}
These equations give renormalized values $C_1(0)$ and $C_2(0)$ 
that agree with the one-loop perturbation theory, namely, with 
the linear terms in the respective right-hand sides of Eqs.~(\ref{mrm}) and (\ref{c2rc2})
(where $C_1=m^2/2$), except for the different normalization used in the appendix. 

Naturally, we can express the constants $K$, $Q$, and $R$
in terms of the initial (bare) values $C_n(\L_0)=\L_0^{(5-n)/2}\,c_n(0),\;n=1,\ldots,4.$
The set of RG trajectories such that $C_1(0)=0$ is given by the condition 
$C_1(\L_0)+8\L_0C_3=0$, according to Eq.~(\ref{RGevol1}). This condition can 
as well be expressed as $K=0$ in Eq.~(\ref{RGline1}). 
In geometric terms, it gives the hyperplane tangent at the origin to the manifold 
such that the renormalized $C_1=m^2/2$ is null, that is to say, 
the maximal critical manifold.
Let us note that  
the right eigenvector $(-8,0,1,0)$ is contained in that hyperplane and is 
tangent to the RG trajectory that leads to the Wilson-Fisher fixed point, 
while keeping $c_2=c_4=0$. 

In analogy with the trajectories that make null the renormalized $C_1$, 
the conditions $C_2(\L_0)+8\L_0C_4=0$ or $Q=0$ define the hyperplane 
that is tangent at the origin to the manifold with null renormalized $C_2$. 
Since we need that both the renormalized $C_1$ and $C_2$ be null 
to have a {\em true} critical point, as shown in Sect.~\ref{RG}, 
we expect that the set of trajectories with $K=Q=0$ 
leads to a fixed point. Of course, this set of trajectories is parametrized by $R$ and 
depending on the value of this parameter, the end fixed point could be different. 
Given that the $S_3$-invariant fixed point has codimension three, it should 
correspond to a definite value of $R$, whereas the $O(2)$-invariant fixed point, 
with codimension two, should correspond to an interval of $R$, with the point $R=0$ 
in the middle. 
Nothing more can be said without an analysis beyond the linear ERG.

Somehow, we are assuming a simple type of RG flow, that is to say, 
a gradient flow or some slight generalization of it, for example, 
a Morse-Smale flow \cite{Gu-Ho}. 
In two dimensions, flows are subject to topological restrictions, as implied by 
the Poincar\'e-Bendixson theorem \cite{Gu-Ho}. 
The examples studied by Nicoll et al \cite{N-C-Stanley_1} show that  
there can be simple RG flows in higher dimensions. 
However, the possibility of very complex RG flows 
has to be borne in mind \cite{Morozov}. Nevertheless, 
advanced mathematical methods could still be applied; 
for example, the Conley theory \cite{Gukov}.

\section{Summary and conclusion}
\label{conclu}

We study statistical models with $S_3$ symmetry that generalize the 3-state Potts model, 
employing the local potential approximation of the sharp-cutoff formulation of the non-
perturbative renormalization group.  We look for a scale-invariant potential, that is to say, 
a fixed point of the partial differential equation ruling the flow of the potential.  While 
the non-trivial fixed point solution of the partial differential equation can be found for 
the isotropic vector model, the equation for models with $S_3$ symmetry is more difficult and 
one resorts to truncated potentials, defined by a limited number of coupling constants.  
Thus, the partial differential equation boils down to a set of ordinary differential 
equations (the beta-functions of that set of coupling constants), whose fixed points are 
given by the solution of a set of algebraic equations.  

A common problem of the method is that the 
solutions change with the order of truncation. 
Therefore, it is crucial, first, to find 
solutions of the set of algebraic equations for fixed points and, second, 
to determine whether the fixed points are real, 
that is to say, are an approximation of the actual fixed points of the full 
potential, or they are spurious, that is to say, artifacts of the 
truncation. To do so, the standard criterion is based on studying the convergence 
of the results with the order of truncation. 

The beta-functions equations can be written as polynomial equations. 
The complexity of the polynomials 
grows rapidly with the order of truncation. An approach to consider consists in 
applying advanced methods of solving systems of polynomial equations; that is to say, 
methods of finding a complete set of numerical solutions, since algebraic solutions 
are not possible except for very low-degree systems. 
Regretfully, one discovers that the available algorithms soon become 
too demanding and, in addition, produce inaccurate solutions.

Moreover, there are too many solutions and the 
problem of discarding spurious solutions is difficult to handle, 
in absence of clear convergence with truncation order. 
Seeking an alternative criterion, we are led to adopt a combination 
of the ERG with $\ve$-expansion methods as the best one available; namely, we are led 
to assume that the RG fixed points in lower dimensions 
must arise as bifurcations of the trivial solution at a set of higher dimensions. 
In this we follow preceding studies, but we focus on the bifurcation at $d_\mathrm{c}=10/3$, 
which we find most relevant for $S_3$-invariant models close to three dimensions.  

The bifurcation at $d_\mathrm{c}=10/3$ is seen to be of the pitchfork form, 
yielding two equivalent $S_3$ fixed points, to be identified. 
The corresponding $\ve$-expansion is an expansion in $\sqrt{-\ve}$, 
which seems to yield a real fixed-point solution for growing $d$ only. 
However, we are able to study the actual dimension dependence of 
the bifurcation starting at $d_\mathrm{c}=10/3$, and we find that it yields 
a real RG fixed point in $d=3$. In particular, 
the method of solution tracing that we introduce 
allows us to follow the course of the $S_3$ fixed point 
for values of $d$ far from $d_\mathrm{c}$, with truncations up to 18 couplings. 
Thus we find that above $d_\mathrm{c}=10/3$ there is another critical dimension, 
at which the fixed point position turns to decreasing with $d$ and eventually 
reaches $d=3$. 
This behavior is 
preserved as the order of truncation grows, 
further supporting the physical nature of the $S_3$ fixed point.

Last, we have discussed the meaning of the new $S_3$ fixed point, in regard to 
the other two fixed points. In this respect, 
it is important to analyze their relative stability and other properties of the 
RG flow that involve the three fixed points. The relative 
stability of a non-trivial fixed point is determined by the bifurcations that it undergoes  from the trivial fixed point and is confirmed by computing numerically 
the Jacobian matrix eigenvalues for several truncations of the potential. 
We find that the new RG fixed point is thrice unstable and, 
therefore, is less stable than the $O(2)$-invariant Wilson-Fisher fixed point. 
It is plausible that a RG trajectory, within the critical manifold of the latter, 
connects both fixed points, going from 
the former to the latter.
However, we have not been able to confirm it, 
because of the difficult ERG integrations involved. 

We have instead carried out an analysis of the critical manifold near the trivial 
RG fixed point, in $d=3$ and in the restricted space of the four relevant coupling constants. 
In this space, the critical manifold that we seek is bidimensional (a surface), and it 
can be parametrized with one parameter that measures the distance from the origin and another 
transversal to it (the one that we have called $R$). 
The RG trajectory that leads to the $S_3$ fixed point
corresponds to a definite value of $R$ (which we cannot calculate) and, arguably, 
smaller values of $R$ correspond to trajectories that pass near the $S_3$ fixed point and eventually go to the Wilson-Fisher fixed point, approaching the trajectory that connects 
both fixed points. 

The critical manifold around each non-trivial fixed point also is easily found.  
One would like to piece together the results of the local analysis 
around each RG fixed point, but this should be done by means of complicated numerical integrations. 
This global analysis is, 
of course, related to interesting questions about the general nature of RG flows, 
for example, 
whether they can be chaotic or not. Regarding our problem, we can certainly assert that 
it gives rise to complex flows, but the question of proving (or disproving) chaos 
is certainly a difficult one and is beyond the scope of the present work.

{
Some problems derived from dealing with many couplings can be avoided in 
a perturbative expansion in the relevant coupling constants. 
Given that the $\ve$-expansion is not suitable for 
describing the new $S_3$ fixed point in $d=3$, we should employ perturbation theory 
in fixed $d=3$. To wit, we should develop the method sketched 
in the appendix, sect.~\ref{Perturbative}. In such perturbative treatment, 
one might be able to achieve a good precision with just the 
four relevant coupling constants, namely, $C_2$, $C_3$, and $C_4$, in addition to $m$.
A comparison with the results of the perturbative renormalization group 
is left for the future.
}

{
Of course, within the ERG, the best way of obtaining truncation-independent results is 
not to truncate at all, that is to say, to deal with the full effective potential and 
its partial differential equation, Eq.~(\ref{ERG}) [or Eq.~(\ref{LERG})]. 
This is a procedure more faithful to the functional basis of the ERG. This procedure is 
feasible for the isotropic vector model because the fixed.point equation is actually an 
ordinary differential equation. Unfortunately, with anisotropy, 
it is not so and more mathematical sophistication is required. We are not able 
to make progress along this line yet.
}

Another question is the possible continuation of the $S_3$ fixed point
in $d<3$. Perhaps it is possible to trace it down to $d=2$. This would raise the interesting 
question of identifying its critical behavior with that of some know $d=2$ conformal model. 
In Sect.~\ref{other}, we have noticed that 
the second model of the $W_3$-symmetry series 
has three $S_3$-invariant conformal fields that are RG-relevant and
can be identified as powers of the elementary fields that 
match the first three terms of potential (\ref{13pot}).
This $W_3$-model is certainly a candidate for such identification. 
This task is also beyond the scope of the present work.

\appendix
\section{Perturbative $(C_2\phi^3+C_3\phi^4+C_4\phi^5)$ theory} 
\label{Perturbative}

This appendix only considers the single-field case, for simplicity
and for an easy comparison with Refs.~\onlinecite{Blume-Capel,Gracey}. This 
simplification only causes a change of the numerical values of the coefficients in 
the renormalization formulas
and does not affect the arguments in the main text.

\subsection{One-loop perturbative $(C_2\phi^3+C_3\phi^4+C_4\phi^5)_3$ theory} 

The effective potential at one-loop order is calculated with 
the background-field method, in which one finds the effect of field fluctuations 
on a classical (bare) potential $U_\mathrm{clas}$, as done in the case of a potential 
with a different set of couplings in the appendix of Ref.~\onlinecite{ISP}. 
We now put
$$
U_\mathrm{clas}(\phi) = \frac{m^2}{2}\,\phi^2+ C_2\,\phi^3 + C_3\,\phi^4+ C_4\,\phi^5 ,
$$
that is to say, the renormalizable couplings in $d=3$ (including odd powers of the field, 
of course). Strictly speaking, we should also add the sixth power of the field 
(as in Ref.~\onlinecite{ISP}), turning the above super-renormalizable potential into one 
that is just renormalizable. However, such extra term can only have a small effect, like 
in the $(\l\phi^4)_3$ theory.

Taking into account the expression of the one-loop effective potential \cite{ISP}, 
the renormalization is carried out as follows:
\begin{align}
m_\mathrm{r}^2 &= U_\mathrm{eff}''(0) = m^2 -\frac{9 C_2^2}{4 \pi m} +
\frac{6 C_3 \L_0}{\pi ^2}-\frac{3 C_3 m}{\pi }\,,
\label{mrm}\\
{C_\mathrm{r}}_2 &= \frac{U_\mathrm{eff}'''(0)}{3!} = C_2 + 
\frac{9 C_2^3}{8 \pi  m^3}-\frac{9 C_3 C_2}{2 \pi  m}+\frac{5 C_4 \Lambda _0}{\pi ^2}-\frac{5
   C_4 m}{2 \pi }\,,
\label{c2rc2}\\
{C_\mathrm{r}}_3 &= \frac{U_\mathrm{eff}^{(4)}(0)}{4!} = C_3  
-\frac{81 C_2^4}{32 \pi  m^5}+\frac{27 C_3 C_2^2}{4 \pi 
   m^3}-\frac{15 C_4 C_2}{2 \pi  m}-\frac{9 C_3^2}{2 \pi  m}\,,
\label{c3rc3}\\
{C_\mathrm{r}}_4 &= \frac{U_\mathrm{eff}^{(5)}(0)}{5!} = C_4 +
\frac{243 C_2^5}{32 \pi  m^7}-\frac{81 C_3 C_2^3}{4 \pi 
   m^5}+\frac{45 C_4 C_2^2}{4 \pi  m^3}+\frac{27 C_3^2 C_2}{2
   \pi  m^3}-\frac{15 C_3 C_4}{\pi  m} \,. 
\label{c4rc4}
\end{align}
It is not difficult to 
associate the various terms with 
definite Feynman graphs. In particular, the two terms proportional to $\L_0$ (divergent) 
are associated with two bubble graphs with a single vertex insertion, either with four or 
five legs.
 
Solving for $m$, $C_2$, $C_3$ and $C_4$, at one-loop order, 
in Eqs.~(\ref{mrm}), (\ref{c2rc2}), (\ref{c3rc3})  
and (\ref{c4rc4}):
\begin{align}
m^2 &= m_\mathrm{r}^2 +
\frac{9 C_{\text{r2}}^2}{4 \pi  m_r}+\frac{3 C_{\text{r3}}
   m_r}{\pi }-\frac{6 \Lambda _0 C_{\text{r3}}}{\pi ^2}\,,
\label{mmr}\\
C_2 &= 
{C_\mathrm{r}}_2 +
\frac{9 C_{\text{r2}} C_{\text{r3}}}{2 \pi  m_r}-\frac{9
   C_{\text{r2}}^3}{8 \pi  m_r^3}+\frac{5 C_{\text{r4}} m_r}{2
   \pi }-\frac{5 \Lambda _0 C_{\text{r4}}}{\pi ^2}\,,
\label{c2c2r}\\
C_3 &= 
{C_\mathrm{r}}_3  
-\frac{27 C_{\text{r2}}^2 C_{\text{r3}}}{4 \pi  m_r^3}+\frac{15
   C_{\text{r2}} C_{\text{r4}}}{2 \pi  m_r}+\frac{81
   C_{\text{r2}}^4}{32 \pi  m_r^5}+\frac{9 C_{\text{r3}}^2}{2
   \pi  m_r}
\label{c3c3r}\\
C_4 &= 
{C_\mathrm{r}}_4 +
\frac{81 C_{\text{r2}}^3 C_{\text{r3}}}{4 \pi  m_r^5}-\frac{27
   C_{\text{r2}} C_{\text{r3}}^2}{2 \pi  m_r^3}-\frac{45
   C_{\text{r2}}^2 C_{\text{r4}}}{4 \pi  m_r^3}-\frac{243
   C_{\text{r2}}^5}{32 \pi  m_r^7}+\frac{15 C_{\text{r3}}
   C_{\text{r4}}}{\pi  m_r}\,.
\label{c4c4r}
\end{align}
Equations (\ref{mmr}) and (\ref{c2c2r}) express how the divergences in the limit 
$\L_0 \ra\infty$ are absorbed in the respective bare couplings.

\subsection{Perturbative $(C_2\phi^3+C_3\phi^4+C_4\phi^5)$ beta function} 

For the critical dimension $d_c=10/3$, the 
perturbative calculation of the beta-function has to be carried out to a higher loop order 
than for $d_c=4$. 
In fact, a three-loop calculation is necessary \cite{Blume-Capel,Gracey}. 
The calculation of the effective potential is simplified with the use of 
the background-field method, to the extent that only the two Feynman graphs drawn below
are necessary. 

\setlength{\unitlength}{1cm}
\thicklines

\begin{picture}(10,4)
\put(6.5,2){\circle{2}}
\put(5.8,2){\line(4,-1){1.3}}
\put(5.8,2){\line(4,1){1.3}}
\put(2,2){\circle{2.2}}
\qbezier(1.3,2)(2,2.6)(2.7,2)
\qbezier(1.3,2)(2,1.4)(2.7,2)
\end{picture}

The first Feynman graph (of the so-called ``melon'' or ``sunset'' type) 
is relatively simple in position space. 
Employing the integral representation of the $D$-dimensional propagator:
$$
\int d^D\!x \int_0^\infty \frac{d\a_1\cdots d\a_4}{(\a_1\cdots \a_4)^{D/2}}\,
\exp\left[-(\a_1+\cdots+ \a_4)\,m^2 -
\left(\frac{1}{\a_1}+\cdots+ \frac{1}{\a_4}\right)\frac{x^2}{4}\right].
$$
The integral over $x$ is easily done. Naturally, the resulting integral can be divergent. 
It is actually divergent for $D \geq 8/3$. 
It can be regularized by taking derivatives with respect to $m^2$ to expose the divergences. After extracting the divergent terms, we can replace $m^2$ with $U_\mathrm{clas}''(\phi)$ 
and take derivatives of $U_\mathrm{eff}(\phi)$ with respect to $\phi$ to obtain the 
contribution of this Feynman graph to renormalization, as in Ref.~\onlinecite{ISP} 
or the preceding section. 

An alternative procedure to deal with ``melon'' Feynman graphs is based on 
the method of Tuthill et al \cite{Tuthill}. This method combines position 
and Fourier space in the graph integral, which can be written as
$$
\int d^D\!x 
\left[\int \frac{d^D\!k}{(2\pi)^{D}} \frac{e^{i\bm{k}\cdot \bm{x}}}{k^2+m^2}\right]^4 =
\O_D \frac{\G(D/2-1)^4}{4^4 \pi^{2D}}\int_0^\infty \frac{dr}{r^{3D-7}}\,F(m^2r^2),
$$
where $\O_D=2\pi^{D/2}/\G(D/2)$ and the function $F$ is such that $F(0)=1$. Again, 
it can be regularized by taking derivatives with respect to $m^2$, which diminishes 
the exponent $3D-7$ and exposes the divergences near $D = 10/3$ 
(a quadratic divergence and a logarithmic divergence).

The second Feynman graph has a more complicated structure in position space. 
The integral representation of the propagators can still be useful but the resulting 
integral is somewhat unwieldy. Therefore, we turn to a less cumbersome procedure.

Since we are just interested in the expansion in $\ve=10/3-D$, we can calculate 
the divergences that arise, for each graph, 
in dimensional regularization in the massless limit, 
as usual \cite{Blume-Capel,Gracey}. However, this calculation is not suitable 
for taking at the end derivatives with respect to $\phi$, which is not present. 
In consequence, in place of the first Feynman graph, we must consider the graph 
that results from it after a vertex insertion in a leg, 
as in Refs.~\onlinecite{Blume-Capel,Gracey}.
Therefore, we have two graphs with three vertices and five legs. 
Both of them are particular cases of a formula obtained by Drummond \cite{Drum}
that applies to any three-vertex graph. Drummond's idea is to regularize 
the infrared divergences of massless Feynman integrals taking advantage 
of their conformal invariance and computing them in a $D+1$-sphere. The formula is 
\begin{align*}
&\int \frac{d^D\!x_1 \, d^D\!x_2}{|x_1-x_2|^\a\, |x_2-x_3|^\b\, |x_3-x_1|^\gamma} = \\
(2R)^{2D-\a-\b-\gamma} \,
&\frac{\G[D-(\a+\b+\gamma)/2]\,\G[(D-\a)/2] \,\G[(D-\b)/2]\, \G[(D-\gamma)/2]}
{\G[(D-\a-\b)/2] \,\G[(D-\a-\gamma)/2]\, \G[(D-\b-\gamma)/2]},
\end{align*}
where $R$ is the radius of the $D+1$-sphere. The total power of the numerator 
in the integral is $2D$ and of the denominator is $\a+\b+\gamma$. For the 
first graph (with a vertex insertion), $(\a,\b,\gamma)=(1,3,1) \times (D-2)$, 
while for the second graph, $(\a,\b,\gamma)=(2,1,2) \times (D-2)$. In both cases, 
the total power of the denominator is $5D-10$, so that the powers are equal 
if $5D-10=2D$ or $D=10/3$. At this dimension, the integral has a logarithmic divergence, 
which appears as a pole of $\G[D-(\a+\b+\gamma)/2]$.

Of course, we also need the combinatorial factors for the Feynman graphs in 
the background-field expansion. The first Feynman graph has a symmetry factor 
$1/(2 \times 4!) = 1/48$ and the second graph $1/(2\times 2!\times 2!)=1/8$. 
Summing the total contribution of the two graphs and taking the derivative with 
respect to the only scale $R$, we obtain the beta function.  
Note that the derivative with respect to $R$ cancels the pole at $D=10/3$ and yields: 
\begin{equation}
\frac{dC_4}{dR} = \frac{3\ve}{2}\,C_4 + \frac{1377\cdot 5!^2}{16\,(4\pi)^5} \G(2/3)^3\, C_4^3.
\label{betaR}
\end{equation}
This result agrees with Refs.~\onlinecite{Blume-Capel,Gracey}, after the pertinent  redefinition of the coupling constant (it agrees except for a typo in the sign in 
Ref.~\onlinecite{Gracey}).

{\section{Beta-functions} }
\label{betas}

{
Here are displayed a few initial beta-functions with the sharp-cutoff regulator and with 
Litim's regulator, respectively, to facilitate a general comparison. Let us notice that  
the beta-functions for Litim's regulator have denominators that 
are powers of $(1+2c_1)$ of one degree higher. This is the essential difference. 
The numerical coefficients are also different, but we must recall that 
the first set of beta-functions is normalized to remove the numerical constant $A_d$ 
whereas the second set is normalized to remove $B_d=A_d/d$. Thus, to compare 
the numerical coefficients, one should revert those redefinitions.
}

{
Sharp-cutoff beta-functions:
\begin{align}
\b_1 &= 2 c_1-\frac{18 c_2^2}{(2 c_1+1)^2}+\frac{8 c_3}{2
   c_1+1},\\
\b_2 &= \frac{6-d}{2}\, c_2 -\frac{24 c_3 c_2}{(2
   c_1+1)^2}+\frac{8 c_4}{2 c_1+1},\\
\b_3 &= (4-d)\,c_3 -\frac{324
   c_2^4}{(2 c_1+1)^4}+\frac{288 c_3 c_2^2}{(2
   c_1+1)^3}-\frac{8 \left(5 c_3^2+9 c_2 c_4\right)}{(2
   c_1+1)^2}+\frac{9 (c_5+2 c_6)}{2 c_1+1},\\
\b_4 &= \frac{10-3d}{2}\, c_4-\frac{864 c_3 c_2^3}{(2
   c_1+1)^4}+\frac{96 \left(3 c_4 c_2^2+4 c_3^2
   c_2\right)}{(2 c_1+1)^3}-\frac{2 (56 c_3 c_4+63 c_2
   c_5+36 c_2 c_6)}{(2 c_1+1)^2}\nonumber\\ &\phantom{aaa} + \frac{20 c_7}{2 c_1+1}\,.
\end{align}
}

{
Beta-functions with Litim's regularization (see also Ref.~\onlinecite[eq.~3.36]{BAZ}, 
in which a different normalization of coupling contants is used):
\begin{align}
\b_1 &= 2 c_1-\frac{72 c_2^2}{(2 c_1+1)^3}+\frac{16 c_3}{(2
   c_1+1)^2},\\
\b_2 &= \frac{6-d}{2}\, c_2-\frac{96 c_3 c_2}{(2
   c_1+1)^3}+\frac{16 c_4}{(2 c_1+1)^2},\\
\b_3 &= (4-d)\,c_3-\frac{2592 c_2^4}{(2 c_1+1)^5}+\frac{1728 c_3
   c_2^2}{(2 c_1+1)^4}-\frac{32 \left(5 c_3^2+9 c_2
   c_4\right)}{(2 c_1+1)^3}+\frac{18 (c_5+2 c_6)}{(2
   c_1+1)^2},\\
\b_4 &= \frac{10-3d}{2}\, c_4-\frac{6912 c_3 c_2^3}{(2
   c_1+1)^5}+\frac{576 \left(4 c_3^2+3 c_2 c_4\right)
   c_2}{(2 c_1+1)^4}-\frac{8 (56 c_3 c_4+63 c_2 c_5+36
   c_2 c_6)}{(2 c_1+1)^3}\nonumber\\ &\phantom{aaa} +\frac{40 c_7}{(2 c_1+1)^2}\,.
\end{align}
}


\begin{thebibliography}{99}

\bibitem{Wu82} F.-Y. Wu,   
Reviews of Modern Physics 54, 235 (1982).

\bibitem{ZW75} R.K.P. Zia and D.J. Wallace,  
Journal of Physics A: Mathematical and General 8, 1495 (1975).

\bibitem{LL}
L.D. Landau and E.M. Lifshitz, \emph{Statistical Physics, Part 1}, 3rd ed.; 
Pergamon Press: {Oxford, UK} (1980).
ISBN O-O8-O23O39-3

\bibitem{S-Fisher}
J.P. Straley and M.E. Fisher,
Journal of Physics A: Math., Nucl. and Gen. 6, 1310 (1973).

\bibitem{Alex74} S. Alexander, 
Solid State Communications 14, 1069--1071 (1974).

\bibitem{Golner73} G.R. Golner, 
Physical Review B 8, 3419 (1973).

\bibitem{Wil-Kog} K.G. Wilson and J. Kogut, Phys. Rept. 12C, 75 (1974).

\bibitem{Amit76} D.J. Amit,  
Journal of Physics A: Mathematical and General 9, 1441 (1976).

\bibitem{mcrit-LP}
A. Codello, M. Safari, G.P. Vacca, and O. Zanusso,
Phys. Rev. D 102, 125024 (2020).

\bibitem{Newman84} K.E. Newman, E.K. Riedel, and S. Muto,  
Physical Review B 29, 302 (1984).

\bibitem{MOP} A. Margaritis, G. \'Odor and A. Patk\'os, Z. Phys. C39, 109 (1988).

\bibitem{BAZ} R. Ben Ali Zinati, A. Codello, 
J. Stat. Mech. 013206 (2018).

\bibitem{S-V}
C.A. S\'anchez-Villalobos, B. Delamotte, N. Wschebor,
Phys. Rev. E 108, 064120 (2023).

\bibitem{FucitoP} F. Fucito and G. Parisi, J. Phys. A 14, L499 (1981).

\bibitem{Wegner-H} F.J. Wegner and A. Houghton, Phys. Rev. A 8, 401 (1973).

\bibitem{N-C-Stanley}
J.F. Nicoll, T.S. Chang, H.E. Stanley, 
Phys. Rev. Lett. 33, 540 (1974).

\bibitem{N-C-Stanley_1}
J.F. Nicoll, T.S. Chang, H.E. Stanley, 
Phys. Rev. B 12, 458 
(1975).

\bibitem{Polchinski} J. Polchinski, Nucl. Phys. B231, 269 (1984).

\bibitem{Hasen2} A. Hasenfratz and P. Hasenfratz, Nucl. Phys. B270, 687 (1986).

\bibitem{Felder}
G. Felder, 
Comm. Math. Phys. 111, 101 (1987).

\bibitem{Wett} C. Wetterich, 
Phys. Lett. B 301, 90 (1993).

\bibitem{Alford} M. Alford, Phys. Lett. B 336, 237 (1994).

\bibitem{Morris} T. Morris, Int. J. Mod. Phys. A9, 2411 (1994).

\bibitem{Morris_1} T.R. Morris, 
Phys. Lett. B 329, 241 (1994). 

\bibitem{Morris_2} T.R. Morris, 
Phys. Lett. B 334, 355 (1994). 

\bibitem{Wiese}
K.J. Wiese and J.L. Jacobsen, 
J. High Energ. Phys. 2024, 92 (2024).

\bibitem{KopietzBS} P. Kopietz, L. Bartosch, and F. Schutz. Introduction to the functional renormalization group. Lect. Notes Phys., 798:1--380 (2010).

\bibitem{Delamotte} B. Delamotte. An Introduction to the nonperturbative renormalization group. Lect. Notes Phys., 852:49--132 (2012). 

\bibitem{Rosten} O.J. Rosten, 
Physics Reports 511, 177--272 (2012).

\bibitem{Dupuis} N. Dupuis, L. Canet, A. Eichhorn, W. Metzner, J. M. Pawlowski, M. Tissier, and N. Wschebor, 
Physics Reports 910, 1--114 (2021).

\bibitem{Plato} R. Ben Ali Zinati, A. Codello, G. Gori, 
Jour. of High En. Phys., Article nb.: 152 (2019). 

\bibitem{Blume-Capel} 
A. Codello, M. Safari, G.P. Vacca, and O. Zanusso,
Phys. Rev. D 96, 081701 (2017).

\bibitem{Gracey}
J. A. Gracey, 
Eur. Phys. J. C  80:604 (2020).

\bibitem{II} J. Gaite, 
Universe 9, 409 (2023).

\bibitem{Gukov}
S. Gukov,
Nuclear Physics B 919, 583--638 (2017).

\bibitem{Nien-RS}
B. Nienhuis, E.K. Riedel, and M. Schick,
Phys. Rev. B 23, 6055 (1981).

\bibitem{Comellas-T} 
J. Comellas, A. Travesset, Nuclear Physics B 498 [FS] 539--564, (1997).

\bibitem{Jakub} 
P. Jakubczyk, N. Dupuis, and B. Delamotte,
Phys. Rev. E 90, 062105 (2014).

\bibitem{Bezout}
B. Sturmfels, 
\emph{Solving Systems of Polynomial Equations,} 
American Math.\ Society (2023).

\bibitem{Gavai-K} 
R.V. Gavai and F. Karsch,
Phys. Rev. B 46, 944--954 (1992).

\bibitem{Krasnytska}
M. Krasnytska, P. Sarkanych, B. Berche, et al,  
Eur. Phys. J. Spec. Top. 232, 1681--1691 (2023). 

\bibitem{MuTa}
A.A. Mukovnin and V.M. Talanov, 
Solid State Communications 152, 2013--2017 (2012).

\bibitem{Kutin} 
E I Kut'in, V L Lorman and S V Pavlov,
Sov. Phys. Usp. 34, 497 (1991).

\bibitem{I} J. Gaite,
J. Phys. A: Math. Gen. 25, 3051 (1992).

\bibitem{IMM}
J. Gaite, J. Margalef-Roig and S. Miret-Art\'es,
Phys. Rev. B 59, 8593 (1999).

\bibitem{Oshi}
M. Oshikawa,
Phys. Rev. B 61, 3430 (2000).

\bibitem{L-S-Balents}
J. Lou, A.W.\ Sandvik, and L. Balents,
Phys. Rev. Lett. 99, 207203 (2007).

\bibitem{SFT}
C. Itzykson, J. Drouffe, 
\emph{Statistical Field Theory,} 
vol.~2, 
Cambridge U.\ Press, New York (1989).

\bibitem{Zamo}
A.B. Zamolodchikov,
Sov. J. Nucl. Phys. 44, 529 (1986).

\bibitem{Kuby}
Y Kubyshin, R Neves, and R Potting,
Proceedings of the Workshop on the Exact Renormalization Group held in Faro, Portugal, in September 10--12, World Scientific (1998). 

\bibitem{Codello}
A. Codello, 
J. Phys. A: Math. Theor. 45, 465006 (2012).

\bibitem{FatZamo}
V.A. Fateev and A.B. Zamolodchikov,
Nucl. Phys. B 280, 644--660 (1987).

\bibitem{Bal-Den}
S. Balaska and K. Demmouche,
Mod. Phys. Lett. A 19, 2135--2145 (2004).

\bibitem{Muss} G. Mussardo, M. Panero and A. Stampiggi, 
J. Stat. Mech. 033103 (2024).

\bibitem{Jan-Vila}
W. Janke and R. Vilanova, 
Nucl. Phys. B, 679--696 (1997).

\bibitem{Chester-Su}
S.M. Chester and N. Su,
Upper critical dimension of the 3-state Potts model, arXiv:2210.09091.

\bibitem{Koh-Yang}
I.-G. Koh and S.-K. Yang,
Phys. Lett. B 223, 349--352 (1989).

\bibitem{IW}
J. Gaite,
Nucl. Phys. B 411, 321--339 (1994).

\bibitem{IW1}
J. Gaite,
Phys. Lett. B 325, 51--57 (1994).

\bibitem{Fisher}
M.E. Fisher, Phys. Rev. Lett. 40, 1610 (1978).

\bibitem{Gu-Ho}
J. Guckenheimer and P. Holmes, 
\emph{Nonlinear Oscillations, Dynamical Systems, and Bifurcations of Vector Fields.} 
Applied Mathematical Sciences, 42. Springer-Verlag, New York (1983).

\bibitem{Go-Sch}
M. Golubitsky and D.G. Schaeffer, 
\emph{Singularities and Groups in Bifurcation Theory, vol I.} 
Applied Mathematical Sciences, 51. Springer-Verlag, New York (1985).


\bibitem{Aoki}
K.-I. Aoki, K. Morikara, W Souma, J.-I. Sumi and H. Terao,
Progress of Theoretical Physics 95, 409--420 (1996).

\bibitem{harmonic}
F. Wegner, 
Phys. Rev. B 6, 1891 (1972).

\bibitem{tracing}
C.L. Bajaj, C.M. Hoffmann, R.E. Lynch, and J.E.H. Hopcroft,
Computer Aided Geometric Design 5, 285--307 (1988).

\bibitem{Morozov} A. Morozov and A. Niemi,
Nucl. Phys. B666, 336 (2003).

\bibitem{ISP} J. Gaite, SciPost Phys. Core 5, 044 (2022). ArXiv: 2112.02060v3.

\bibitem{Tuthill} 
G.F. Tuthill, J.F. Nicoll, and H.E. Stanley,
Phys. Rev. B 11, 4579, (1975).

\bibitem{Drum}
I.T. Drummond,
Phys. Rev. D 19, 1123 (1979).


\end{thebibliography}
\end{document}